%% file: pp.tex
\documentstyle[aaspp4]{article}
\tighten

\def\qso{PKS~0454+039}
\def\AlII{\hbox{{\rm Al}\kern 0.1em{\sc ii}}}
\def\AlIII{{\hbox{\rm Al}\kern 0.1em{\sc iii}}}
\def\CaII{\hbox{{\rm Ca}\kern 0.1em{\sc ii}}}
\def\CIVwaves{{\rm C}\kern 0.1em{\sc iv}~$\lambda\lambda 1548, 1550$}
\def\CII{\hbox{{\rm C}\kern 0.1em{\sc ii}}}
\def\CIII{\hbox{{\rm C}\kern 0.1em{\sc iii}}}
\def\CIV{\hbox{{\rm C}\kern 0.1em{\sc iv}}}
\def\CV{\hbox{{\rm C}\kern 0.1em{\sc v}}}
\def\NHI{\hbox{$N$({\rm H}\kern 0.1em{\sc i})}} 
\def\HI{\hbox{{\rm H}\kern 0.1em{\sc i}}}
\def\HII{\hbox{{\rm H}\kern 0.1em{\sc ii}}}
\def\Lya{\hbox{{\rm Ly}\kern 0.1em$\alpha$}}
\def\Lyb{\hbox{{\rm Ly}\kern 0.1em$\beta$}}
\def\Lyg{\hbox{{\rm Ly}\kern 0.1em$\gamma$}}
\def\Lyfive{\hbox{{\rm Ly}\kern 0.1em$5$}}
\def\Lysix{\hbox{{\rm Ly}\kern 0.1em$6$}}
\def\Lyseven{\hbox{{\rm Ly}\kern 0.1em$7$}}
\def\Lyeight{\hbox{{\rm Ly}\kern 0.1em$8$}}
\def\Lynine{\hbox{{\rm Ly}\kern 0.1em$9$}}
\def\Lyten{\hbox{{\rm Ly}\kern 0.1em$10$}}
\def\HeI{\hbox{{\rm He}\kern 0.1em{\sc i}}}
\def\HeII{\hbox{{\rm He}\kern 0.1em{\sc ii}}}
\def\FeI{\hbox{{\rm Fe}\kern 0.1em{\sc i}}}
\def\FeII{\hbox{{\rm Fe}\kern 0.1em{\sc ii}}}
\def\FeIII{\hbox{{\rm Fe}\kern 0.1em{\sc iii}}}
\def\MnII{\hbox{{\rm Mn}\kern 0.1em{\sc ii}}}
\def\MgIIwaves{{\rm Mg}\kern 0.1em{\sc ii}~$\lambda\lambda 2976, 2803$}
\def\MgI{\hbox{{\rm Mg}\kern 0.1em{\sc i}}}
\def\MgII{\hbox{{\rm Mg}\kern 0.1em{\sc ii}}}
\def\MgIII{\hbox{{\rm Mg}\kern 0.1em{\sc iii}}}
\def\NVwaves{{\rm N}\kern 0.1em{\sc v}~$\lambda\lambda 1238, 1242$}  
\def\NV{\hbox{{\rm N}\kern 0.1em{\sc v}}}
\def\NII{\hbox{{\rm N}\kern 0.1em{\sc ii}}}
\def\NIII{\hbox{{\rm N}\kern 0.1em{\sc iii}}}
\def\OVIwaves{{\rm O}\kern 0.1em{\sc vi}~$\lambda\lambda 1031, 1037$} 
\def\OVI{\hbox{{\rm O}\kern 0.1em{\sc vi}}}
\def\OII{\hbox{[{\rm O}\kern 0.1em{\sc ii}]}}
\def\SiIVwaves{{\rm Si}\kern 0.1em{\sc iv}~$\lambda\lambda1393, 1402$}  
\def\SiII{\hbox{{\rm Si}\kern 0.1em{\sc ii}}}
\def\SiIII{\hbox{{\rm Si}\kern 0.1em{\sc iii}}}
\def\SiIV{\hbox{{\rm Si}\kern 0.1em{\sc iv}}}
\def\SII{\hbox{{\rm S}\kern 0.1em{\sc ii}}}
\def\SIII{\hbox{{\rm S}\kern 0.1em{\sc iii}}}
\def\SIV{\hbox{{\rm S}\kern 0.1em{\sc iv}}}
\def\NaI{\hbox{{\rm Na}\kern 0.1em{\sc i}}}
\def\kms{\hbox{km~s$^{-1}$}}       
\def\cm2{\hbox{cm$^{-2}$}}
\def\etal{et~al.\ }
\def\minfit{\sc minfit}
\def\DR{\hbox{\sc dr}}

\slugcomment{Accepted: {\it Astrophysical Journal}}

\begin{document}

\title{High Metallicity Mg II Absorbers in the $z<1$ Ly$\alpha$ Forest \\ 
of {\qso}: Giant LSB Galaxies?\altaffilmark{1,2}}

\thispagestyle{empty}
 
\author{Christopher W. Churchill\altaffilmark{3}}
\affil{Department of Astronomy and Astrophysics \\ 
      The Pennsylvania State University, University Park PA 16802 \\ 
      {\it cwc@astro.psu.edu}}

\and
 
\author{Vincent Le~Brun}
\affil{Laboratoire d' Astronomie Spatiale du CNRS \\ 
       B.P.~8, F--13376, Marseille Cedex 12, France \\
       {\it vlebrun@astrsp-mrs.fr}}

\altaffiltext{1}{Based in part on observations obtained at the
W.~M. Keck Observatory, which is jointly operated by the University of
California and the California Institute of Technology.}
\altaffiltext{2}{Based in part on observations obtained with the
NASA/ESA {\it Hubble Space Telescope}, which is operated by the STSci
for the Association of Universities for Research in Astronomy, Inc.,
under NASA contract NAS5--26555.}
\altaffiltext{3}{Visiting Astronomer at the W. M. Keck Observatory}

%%%%%%%%%%%%%%%%%%%%%%%%%%% ABSTRACT %%%%%%%%%%%%%%%%%%%%%%%%%%%%%%%%%%%%%%%%%%

\begin{abstract}
We report the discovery of two iron--group enhanced high--metallicity
{\MgII} absorbers in a search through 28 {\Lya} forest clouds along
the {\qso} sight line.  
Based upon our survey and the measured redshift number densities of
$W_{\rm r}({\MgII}) \leq 0.3$~{\AA} absorbers and {\Lya} absorbers at
$z \sim 1$, we suggest that roughly 5\% of {\Lya} absorbers at $z\leq
1$ will exhibit ``weak'' {\MgII} absorption to a 5$\sigma$ $W_{\rm
r}(\lambda 2796)$ detection limit of 0.02~{\AA}.
The two discovered absorbers, at redshifts $z=0.6248$ and $z=0.9315$,
have $W_{\rm r}({\Lya}) = 0.33$ and 0.15~{\AA}, respectively.
Based upon photoionization modeling, the {\HI} column densities are
inferred to be in the range $15.8 \leq \log N({\HI}) \leq
16.8$~{\cm2}.
For the $z=0.6428$ absorber, if the abundance pattern is solar, then
the cloud has $[\hbox{Fe/H}] > -1$; if its gas--phase abundance
follows that of depleted clouds in our Galaxy, then $[\hbox{Fe/H}] >
0$ is inferred.
For the $z=0.9315$ absorber, the metallicity is $[\hbox{Fe/H}] > 0$,
whether the abundance pattern is solar or suffers depletion.
Imaging and spectroscopic studies of the {\qso} field reveal no
candidate luminous objects at these redshifts.
We discuss the possibility that these {\MgII} absorbers may arise
in the class of ``giant'' low surface brightness galaxies,
which have $[\hbox{Fe/H}] \geq -1$, and even $[\hbox{Fe/H}] \geq 0$,
in their extended disks.
% (McGaugh 1994, ApJ, 426, 135; Pickering \& Impey 1995, BAAS, 186, 39.07).
We tentatively suggest that a substantial fraction of these ``weak''
{\MgII} absorbers may select low surface brightness galaxies out to $z
\sim 1$.
\end{abstract}


%%%%%%%%%%%%%%%%%%%%%%%%%%% SECTION %%%%%%%%%%%%%%%%%%%%%%%%%%%%%%%%%%%%%%%%%%%

\section{Introduction}
\label{sec:intro}

Metal--line absorption in the intergalactic medium, or
IGM,\footnote{Throughout this paper, we use the terms ``{\Lya}
cloud'', ``forest cloud'', and ``IGM'' somewhat interchangeably to
designate {\Lya} absorption with $\tau_{912} < 1$.} is
astrophysically interesting because the absorption properties can be
exploited to reveal the star formation, chemical enrichment, and
ionization histories of the universe.
This provides a motivation for studying metal lines in {\Lya} absorbers
over as wide a redshift range as possible and for sampling transitions
covering as many ionization levels and chemical species as possible
(cf.~{\cite{uffe}; \cite{rauch97}).
Observations (\cite{hulya}; \cite{lulya}; \cite{kim}) and numerical
simulations (\cite{jordi}; \cite{zhang}; \cite{rad}; \cite{norman})
have revealed that the forest is rapidly evolving with
redshift from $z\sim4$ to $z\sim1$, and that the absorbing gas is
housed in a wide range of cosmic structures undergoing a wide range of
dynamical processes. 
At $z\sim 2$, {\Lya} clouds contain the majority of the baryon content
of the universe.
At lower redshifts, {\Lya} clouds are thought to be more directly
associated with low surface brightness and/or dwarf galaxies
(\cite{salpeter}; \cite{shull96}; \cite{linder}), with the outer disks
and halos of high surface brightness galaxies (\cite{lanzetta};
\cite{lebrunlya}), or with the remnant material left over from the
formation of galaxies and/or small galaxy groups (\cite{vangorkem};
\cite{bowen}; \cite{lebrunlya}).
Studies of the metal content and ionization conditions in these low
redshift forest clouds, especially in the context of their association
(or lack of association) with galaxies, could provide the
``missing--link'' evidence necessary for inferring the evolving
interplay between the IGM and galaxies or the presence of low surface
brightness galaxies at higher redshifts.

A limited number of strong metal--line species have now been
seen in high ionization transitions at $z\sim 2.5$ (\cite{tytlereso};
\cite{cowie}; \cite{songaila}).
However, the chemical and ionization conditions of {\Lya} clouds at
low redshifts ($z\leq1$) remain unexplored because they require
time--intensive programs using HST.  
Relative to $z\sim2.5$, the meta--galactic UV background flux (UVB) at
$z<1$ is reduced by a factor of $\sim 5$ and its shape may be softened
by stellar photons escaping bright field galaxies
(\cite{jeanmichel97}; \cite{giallongo97}; \cite{bergeronkp94}, and
references therein).
Thus, the IGM ionization conditions may have evolved so that low
ionization species, especially the resonant {\MgIIwaves} doublet and
several of the stronger {\FeII} transitions, are detectable  in
{\Lya} clouds.  
As discussed below, these particular species are well suited for
understanding chemical enrichment histories.

Songaila \& Cowie (1996, hereafter SC96) detected {\CIV} absorption in
$\simeq 75$\% of all {\Lya} clouds at $z\sim2.5$ with $\log N({\HI}) \geq
14.5$~{\cm2} and concluded that roughly 50\% of $\log N({\HI}) \leq
14.5$~{\cm2} clouds could have primordial abundances.
They also reported {\SiIV} and {\CII} absorption in a fraction of the
{\Lya} clouds (including ``partial'' Lyman limit systems).
Based upon the photoionization models of Bergeron \& Stasi\'nska
(1986)\nocite{jbands}, SC96 find the metallicity of $z\sim2.5$ {\Lya}
clouds to be $[Z/Z_{\odot}] \sim -2$ and to be fairly uniform, with
about 1~dex of scatter.\footnote{Throughout this paper, we use the
notation $[Z/Z{\odot}] = \log Z - \log Z_{\odot}$, and $[X/Y] = \log
(X/Y) - \log (X/Y)_{\odot}$, where $X$ and $Y$ are any two elements.}
They also report [Si/C] ratios consistent with Galactic Halo stars
(metal poor late--type stars), in that the $\alpha$--group silicon is
enhanced by a factor of three over the carbon.
This conclusion, however, is sensitive to the assumed UVB continuum
shape, especially the question of how much star bursting galaxies
contribute to the UVB, and its non--uniformity, at higher redshifts
(\cite{giroux}).

Considering the mechanisms and range of environments that could
plausibly give rise to metals in what are traditionally known as
{\Lya} clouds, it is difficult to understand a high level of
uniformity in their chemical enrichment histories.
As proposed by Tytler \etal (1995)\nocite{tytlereso}, there are at
least three obvious mechanisms for the enrichment.

(1) The larger $N({\HI})$ clouds may be gravitationally bound with
internal gravitational instabilities in which they produce their
own stars, which in turn distribute the metals throughout the cloud.
This type of object has little distinction from a galaxy.
This {\it in situ\/} process would likely give rise to a strong
metallicity dependency with {\HI} column density, unless a well--tuned
mechanism governing star formation yielded a uniform chemical
enrichment history of the IGM, as suggested by Cowie \etal (1995).
Such a mechanism would likely represent non--standard star formation
processes.

(2) The metals may be produced in protogalaxies and then be
widely distributed via mechanical ejection from merging events
(\cite{gnedin96}) or from correlated supernovae (SNe) (\cite{cen92}).  
This implies that {\Lya} clouds formed after the metals were
distributed around the metal producing galaxies.  
The scenario also predicts that the metal enriched {\Lya} clouds, as
opposed to ``{\Lya}--only'' clouds, would cluster like galaxies.

(3) Population III stars, formed at $z> 10$ and somewhat uniformly
spread throughout the IGM, may have distributed metals into the IGM
prior to the first protogalaxies.
A population of {\Lya} forest clouds that have been enriched by
Population III stars may exhibit IGM chemical conditions that are
relatively unchanged from the epoch of the first stars.
If so, this population would be ideal for studying the intensity and
continuum shape evolution of the UVB from $0<z<4$, since the
changing ionization conditions could be used to deconvolve the
non--evolving chemical conditions from the evolving UVB.
As such, the detection of extremely metal poor stars in the Galaxy
halo would also be very interesting, since their presence would
suggest the presence of Population III stars (cf.~\cite{ostriker}).

The chemical abundance pattern can serve as a clue to the origin,
physical environment, and chemical enrichment history of any given
{\Lya} absorber.
There are at least two major uncertainties involved in measuring 
relative abundances using QSO absorption lines: ionization
corrections and dust grain depletion patterns.  
The ionization correction provides good reason for observing a wide
range of ionization levels.
Moreover, the ionization corrections are sensitive to the intensity
and continuum shape of the ionizing radiation, possibly providing even
further leverage for understanding local environments and chemical
enrichment history.
Dust depletion does not effect all $\alpha$--group elements (for
example, sulfur is not readily incorporated onto dust grains) nor all
Fe--group elements (for example, zinc).  
However, those elements provide neither the strong UV absorption lines
needed to accurately probe clouds with $\tau_{912} < 1$, nor the
transitions observable from the ground for redshifts below $z\sim 1$.
The strongest observable transitions are the {\MgIIwaves} doublet
($\alpha$--group), and {\FeII} $\lambda 2344$, 2382, and 2600.
Unfortunately, both magnesium and iron can deplete onto dust grains
and their depletion levels are environment dependent.

For each of the three scenarios suggested above, it is expected that
the relative elemental abundances of an enriched cloud should reflect
the $\alpha$--group enhanced yield of Type II SNe (note that this is
consistent with the results of SC96).
In essence, the picture is simple: if the $\alpha$--group elements are
enhanced relative to the Fe--group then the chemical enrichment has
been dominated by Type II SNe.  
Based upon the $\hbox{[Si/Fe]}$, $\hbox{[S/Fe]}$, $\hbox{[O/Fe]}$
ratios measured in Galactic halo stars, this pattern is seen for
$\hbox{[Fe/H]} \leq -1$ (see \cite{lauroesch}).
In the case of scenario (3) presented above, it is likely that only a
single burst, or episode, of star formation would have occurred and
that the metal production arise exclusively from Type II SNe.  
In the case of scenario (2), the {\Lya} clouds would be far from the
galaxies; the only metal enriched gas that could infiltrate clouds
forming from the primordial IGM would necessarily be ejected from
correlated Type II SNe bursts.

If the abundance pattern is more in line with solar proportions
(i.e,~$\hbox{[Mg/Fe]} \sim \hbox{[Si/Fe]} \sim \hbox{[Fe/H]} \sim 0$),
then the picture is that Fe--group elements have been built up over a
longer time scale via Type Ia SNe.
This implies a star formation history local to the cloud that would
have been relatively quiescent for $\sim \hbox{Gyr}$ prior to the
epoch of the observed absorption.
Thus, if a given {\Lya} cloud is measured to have $\hbox{[Fe/H]} \geq
-1$ and $\hbox{[$\alpha$/Fe]}$--group abundance ratios near solar
proportions, then one might infer that Type Ia SNe have played a role
in the cloud's chemical enrichment history.
However, based upon uncertainties in Type II SNe yields, Gibson,
Loewenstein, \& Mushotzky (1997\nocite{gibson}) have cautioned that 
the relative importance of Type Ia and Type II SNe as inter--cluster
polluters remains uncertain.

A type of extended gas--rich object that is seen to have
$\hbox{[Fe/H]} \geq -1$, and even $\hbox{[Fe/H]} \geq 0$, is the class
of giant low surface brightness galaxies (\cite{bothun};
\cite{pickering95}; \cite{mcgaugh}).
At low redshifts, the general population of low surface brightness
galaxies are seen to outnumber high surface brightness galaxies by a
factor of at least two (\cite{dalcanton}).
As such, these galaxies are important tracers of low density dark
matter halos and structure formation from small over--density
fluctuations. 
They also may represent environments where the pathways of star
formation and chemical evolution reflect non--standard
astrophysical processes (Bothun \etal 1997\nocite{bothun}).
If these objects are observable in absorption over a wide range of
redshifts, they likely will provide us a unique astrophysical
laboratory for broadening our present perspective on star and galaxy
formation.

In this paper, we report the search for and discovery of {\FeII} and
{\MgII} doublet absorption in the {\Lya} forest along {\qso} sight
line.
In \S\ref{sec:data}, we describe the data and analysis.
In \S\ref{sec:search}, we describe the sample of {\Lya} lines and
our search method.
The absorption properties of the detected systems are presented in
\S\ref{sec:systems}.
We apply photoionization models to the detected metal--line systems in
\S\ref{sec:models}, and briefly discuss the model results in
\S\ref{sec:results}.
The implications of the results are addressed in
\S\ref{sec:discuss}.
A brief summary is provided in \S\ref{sec:conclude}.

%%%%%%%%%%%%%%%%%%%%%%%%%%% SECTION %%%%%%%%%%%%%%%%%%%%%%%%%%%%%%%%%%%%%%%%%%%

\section{Observations and Data Analysis}
\label{sec:data}

The {\MgII} and {\FeII} transitions were searched for in an $R=45,000$
optical HIRES (\cite{vogtspie}) spectrum.
The {\Lya} line list was obtained from the $R=1300$ G190H and G270H
FOS/HST spectra of Boiss\'e \etal (1997, hereafter BBLD).
Three images of the {\qso} field have been incorporated into our
study so that we may attempt to identify the luminous objects giving
rise to the absorption.
Two are high--spatial resolution WFPC2/HST images using the F450W and
F702W filters (Le~Brun \etal 1997, hereafter LBBD).
The third is the deep $R[6930/1500]$ image (centered on $\lambda 6930$
with a FWHM band pass of 1500~{\AA}) taken from Steidel \etal
(1995\nocite{ccs0454}). 
We also draw upon the published (\cite{ccs0454}) and
unpublished (\cite{chuckprivcomm}) spectroscopic redshift measurements
of the many objects along the line of sight to the QSO.

The HIRES spectrum was obtained and reduced as described in
Churchill (1997a\nocite{mythesis}) and in Churchill, Vogt, \& Charlton
(1998\nocite{cvc98}).
The HIRES spectrum has wavelength coverage $3767-6198$~{\AA}, though
there are some breaks in the coverage redward of 5100~{\AA} because
the single setting of the 2048X2048 CCD did not capture the complete
free spectral range at these wavelengths.
The FOS spectra were obtained, reduced, and the list of
{\Lya} forest lines used for this study were produced as described in
BBLD.
The acquisition and analysis of the high--spatial resolution WFPC2
images are described in LBBD.
They also present a synopsis of candidate galaxies along the line of
sight to the quasar available from their study and the literature.
The $R$--band ground--based image and the spectroscopic identifications
of objects in the {\qso} field are described in Steidel \etal
(1995).

%%%%%%%%%%%%%%%%%%%%%%%%%%% SECTION %%%%%%%%%%%%%%%%%%%%%%%%%%%%%%%%%%%%%%%%%%%

\section{Searching the Forest}
\label{sec:search}
 
We searched the HIRES/Keck spectrum of {\qso} for {\MgII} absorption
in the {\Lya} lines reported by BBLD.
In the FOS spectra, the redshift range over which {\Lya} was detected
is $0.4163 \leq z(\lambda 1216) \leq 1.3431$.
The redshift range over which the {\Lya} $\lambda 1215$ transition
could have been detected was $0.41 \leq z \leq 1.69$.
The line list is presented in Table~\ref{tab:mgiilimits}. 
In all cases, they are {\Lya}--only systems (no other corroborating
transitions are detected in absorption).
We have included only those lines from BBLD that are not
confused by blending or are not coincident in wavelength with 
strong metal--line transitions from the four known metal--line
systems ($z = 0.072, 0.859, 1.068, 1.153$) along the line of sight.
The first two columns of Table~\ref{tab:mgiilimits} are the redshift of
{\Lya} absorption and the rest--frame $\lambda 1215.67$ equivalent
width, respectively.
In columns 3--5, values of the neutral hydrogen column density,
$\log N({\HI})$, are tabulated for Doppler $b$ values of 80, 30, and
15~{\kms}, respectively.
These are shown only to illustrate the plausible $N({\HI})$ range that
might be inferred from the equivalent widths.
A $b=30$~{\kms} is representative of the median $b$ value of 34~{\kms}
found by Kim \etal (1997\nocite{kim}) at redshifts $2<z<3$.  
There is evidence that the median $b$ value increases with decreasing
redshift, which is believed to be due to kinematic substructure
evolution for $N({\HI}) \leq 14.0$~{\cm2} clouds. 
For the higher column density clouds, the widths likely reflect
non--shock heated cloud temperatures (\cite{haehnelt96}). %see kim
A $b=80$~{\kms} is plausible for kinematically broadened and/or
shock heated clouds (\cite{kim}).
Doppler parameters greater than $\sim 80$~{\kms} are likely due to
blending (\cite{lulya}), possibly of physically distinct systems.
The $b=15$~{\kms} value is the ``lower cut off'' value found at high
redshift (\cite{lulya}, but also see \cite{hulya}).

The sensitivity of the search, as a function of redshift, is
quantified in terms of the rest--frame limiting equivalent width 
of the {\MgII} $\lambda 2796$ transition.
In Figure~\ref{fig:ewlimits}, we present the sensitivity curve, where
we have chosen to use a 5$\sigma$ significance level.
The redshift range over which {\MgII} doublets could be detected in
the HIRES spectrum is $0.3466 \leq z(\lambda 2796) \leq 1.2134$.
There are small gaps in the coverage above $z(\lambda 2796)=0.8340$
that increase with increasing redshift.
The 5$\sigma$ observed equivalent width detection limit ranged from
0.007 to 0.020~{\AA}, except for $z(\lambda 2796) \leq 0.4662$, where
it ranges from 0.020 to 0.035~{\AA}.
The results, including the detection limits and the limiting column
densities of {\MgII}, which are based upon linear curve of growth
analysis, are presented in columns 6 and 7 of
Table~\ref{tab:mgiilimits}.
Only those redshifts for which both transitions of the {\MgII} doublet
could be observed are tabulated.

Absorption features were defined using our own interactive software,
which is based upon the detection algorithms of the QSO Absorption Line
Key Project (\cite{donspaper}).
The criteria that define a confirmed {\MgII} doublet are presented by 
Churchill \etal (1997)\nocite{crcv97}, who have searched 26 HIRES/Keck
QSO spectra for weak {\MgII} systems.
These lines are fit with Gaussians to obtain their equivalent widths
and observed central wavelengths.
To locate candidate {\MgII} doublets, the candidate $\lambda 2796$
line centroid and detection aperture (full width at the continuum) is
shifted to the expected location of the $\lambda 2803$ line and the
significance level is measured.
An acceptable candidate for the weaker member of the doublet occurs
when the detection significance level is greater than or equal to the
ratio of the transition $f\lambda$ times the significance level of the
stronger member.
Then, a ``chance probability'' is computed by scanning the spectrum
with the detection aperture over the spectrum for $\sim 50$~{\AA} to
both sides of the candidate and computing the fraction of detected
features (both emission and absorption) with a significance level
greater than or equal to the candidate $\lambda 2803$ line.
Most bonafide {\MgII} doublets have chance probabilities of $\leq
10^{-6}$, though a very few have probabilities as large as $\sim
10^{-3}$.

To a 5$\sigma$ limit of $\log N({\MgII}) \sim 11.3$~{\cm2}, the
{\MgII} doublet was detected in two of the twenty--eight {\Lya} lines
in the list, which is a success rate of $\sim 7$\%.
The two weak {\MgII} systems found in the HIRES spectra, have
$z=0.6428$ and $z=0.9315$.
The data for these systems are presented in Figure~\ref{fig:data} and their
measured properties are listed in Tables~\ref{tab:systems} and
\ref{tab:limits}.
For both, the {\MgII} doublet and at least one transition of {\FeII}
was detected.  
Below, we describe the measured properties of the two detected systems.

%%%%%%%%%%%%%%%%%%%%%%%%%%% SECTION %%%%%%%%%%%%%%%%%%%%%%%%%%%%%%%%%%%%%%%%%%%

\section{Properties of the Absorbers}
\label{sec:systems}

In this work, we concentrate on the properties of the two absorbers
for which {\MgII} has been detected.
Here we note that the {\Lya} equivalent widths are among the smallest
in the sample of 28 (there are seven as small as or smaller than the
$z=0.6428$ absorber and two as small as or smaller than the $z=0.9315$
absorber).
Also, we note that the {\MgII} and {\FeII} transitions have been detected
a factor of five to ten above the $5\sigma$ detection limits of the
HIRES spectra.  
Given the stringent limits on the {\MgII} column densities for the
remaining {\Lya} absorbers, and the fact that majority appear to have
higher $N({\HI})$ than the two exhibiting {\MgII}, it may be that there
is a large dynamic range in the $N({\MgII})/N({\HI})$ ratio.
As noted in \S\ref{sec:intro}, this is not the case for {\CIV}
absorption in the {\Lya} forest.

However, we note that upper limits on the metallicities are not very
restrictive if we assume a typical $b$ parameter of 30~{\kms} and 
photoionization by the UVB (using CLOUDY; see Figure~11 of
Churchill \etal 1997\nocite{crcv97}).
For a {\Lya} cloud with $W_{\rm r}({\Lya}) \sim 0.6$~{\AA} and
$N({\MgII}) < 11.4$~{\cm2}, the upper limit on $[Z/Z_{\odot}]$ ranges
from $-0.5$ to $-2.8$, depending upon ionization level.
For a {\Lya} cloud with $W_{\rm r}({\Lya}) \sim 0.4$ and
$N({\MgII}) < 11.4$~{\cm2}, the upper limit ranges from $+0.4$ to $-2.2$
as the cloud becomes highly ionized.
Thus, not a great deal can be said about the range of metallicities in
the {\Lya} forest at $z\leq 1$ based upon our {\MgII} upper limits.

\subsection{The $z=0.6428$ System}

In the left hand panels of Figure~\ref{fig:data}, the {\MgII} and
{\FeII} HIRES profiles are presented.  
Also shown (top panel) is the FOS spectrum, with the position of the
corresponding {\Lya} line marked with a tick.
Along with the {\MgII} doublet, {\FeII} $\lambda2383$ and $\lambda
2600$ were clearly detected.
The weaker {\FeII} transitions were covered by the spectrum, but were
not found to the 5$\sigma$ significance level.  
In part, this is due to the decreasing signal to noise below
4000~{\AA} where the HIRES sensitivity drops rapidly (see
Figure~\ref{fig:ewlimits}).
In the upper panel of Table~\ref{tab:systems}, the measured
rest--frame equivalent widths, column densities and Doppler $b$
parameters are presented.  

The {\MgII} doublet ratio is $1.4\pm0.1$, and both the {\MgII} and
{\FeII} lines are partially resolved.
Based upon the apparent optical depth profiles (cf.~\cite{savage91}),
there is evidence for unresolved saturation in the {\MgII} doublet.
It may be that there are two or more very narrow absorbing components
giving rise the profile, but the signal--to--noise ratio is not high
enough to model the data to this level.
The column densities and $b$ parameters are obtained using Voigt Profile
(VP) fits that incorporated both the atomic physics and the
instrumental spread function.
We used the program {\minfit} (\cite{mythesis}), which performs an
iterative $\chi^{2}$ minimization between the data and the model
spectra [see Churchill (1997a\nocite{mythesis}) for a detailed
description of the convergence criteria and the error computations].
A VP model with two components was handed to {\sc minfit}, but it
returned a single component model based upon the criterion that there
was no statistically significant difference between the best fit
single component and two component models.
The measured VP column densities for the HIRES profiles are $\log
N({\MgII}) = 12.74\pm0.02$~{\cm2} and $\log N({\FeII}) =
12.46\pm0.06$~{\cm2}.
Their respective $b$ parameters are $b_{\rm tot}({\MgII}) =
5.7\pm0.3$~{\kms} and $b_{\rm tot}({\FeII}) = 4.3\pm1.3$~{\kms}. 
The reduced $\chi^{2}_{\nu}$ for the simultaneous VP fit to the
{\MgII} doublet and the two {\FeII} transitions is 0.96, where the
degrees of freedom is $\nu = 107$.
In principle, the contribution of turbulent broadening to the
profiles, $b_{\rm turb}/b_{\rm tot}$, could be determined from the
ratio of the atomic masses of iron and magnesium (see
eq.~[\ref{eq:btot}] and eq.~[\ref{eq:bXofbtot}]), but the
uncertainties are too large to directly place useful limits on $b_{\rm
turb}$.

As we will discuss below, well determined uncertainties in the VP
quantities are central to constraining the ionization, thermal, and
chemical conditions in the absorbing gas ``cloud''.
VP fits are particularly robust for profiles in this regime of column
density and width.
As shown in Churchill (1997a\nocite{mythesis}), the quoted
uncertainties in the VP quantities are consistent with the spread in
these quantities measured from VP fits to 1000 simulated spectra with
similar signal to noise ratios.
Thus, the measured column density and line broadening, and their
uncertainties, are considered to be robust.

In the FOS spectrum, the {\Lya} absorption is the central line in a
triple--blend feature.
In Figure~\ref{fig:midas}, we show the deblending fit. 
The adjacent lines in the blended feature are {\Lyseven} at $z=1.1536$,
and {\Lya} at $z=0.6448$.  
The lone feature at 2004.5~{\AA} is {\Lysix} at $z=1.1536$, which is
the redshift of a strong {\MgII} absorber studied elsewhere
(Churchill \etal 1998\nocite{cvc98}).
Though the fit is not unique, we have adopted the presented result,
which yielded a rest--frame {\Lya} equivalent width of $W_{\rm r} =
0.33\pm0.03$~{\AA}.
Because of the low resolution of the FOS spectrum, we could not obtain
estimates of the {\HI} column density and $b$ parameter directly from
the data.  
There is no coverage at the expected position of the Lyman limit, so
we cannot place an observed upper limit on $N({\HI})$.
The {\Lyb} falls just blueward of the Lyman limit at 1695~{\AA} due to
the $z=0.8596$ damped {\Lya} absorber, and thus cannot be detected.
If the {\HI} and the {\MgII} arise in the same physical locations in
the absorber, then the total {\HI} $b$ parameter is constrained to be
$5.7 \leq b({\HI}) \leq 29.8$~{\kms} (see Figure~\ref{fig:z06428}a)
including the spread introduced by the uncertainty in the measured
$b_{\rm tot}({\MgII})$.
The lower limit corresponds to the case in which turbulence or bulk
motions dominate the line broadening and the upper limit corresponds
to a thermal motions scaling, $\sqrt{24} b_{\rm tot}({\MgII})$.
From the curve of growth, we have estimated that the inferred 
$b({\HI})$ range translates to a neutral hydrogen column density range
of $14.2 \leq \log N({\HI}) \leq 17.6$~{\cm2}.
For this estimate, we have included the spread introduced by the
uncertainty in the measured $W_{\rm r}({\Lya})$, which dominates over
the uncertainty in $b_{\rm tot}({\HI})$.

The FOS spectrum also covers several other transitions from a variety
of species and over a wide range of ionization potentials.  
It is important to thoroughly check for the presence of absorption from
these species and to place limits on their column densities when no
absorption is detected.
These limits may be useful for further
constraining the chemical and ionization conditions of the absorber,
even if the sensitivity level of the FOS spectrum is not very high.
Thus, we have systematically searched the FOS spectrum for other
transitions associated with the $z=0.6428$ absorber, using the
detection method described by Schneider \etal (1993).
Details of the search are presented in Appendix~\ref{app:fossearch}
and selected results are tabulated in Table~\ref{tab:limits}.

\subsection{The $z=0.9315$ System}

In the right hand panels of Figure~\ref{fig:data}, the {\MgII} and
{\FeII} HIRES profiles are presented.  
Also shown (top panel) is the FOS spectrum, with the position of
the corresponding {\Lya} line marked with a tick.
Along with the {\MgII} doublet, the {\FeII} $\lambda2383$ transition
was detected. 
The {\FeII} $\lambda2600$ transition may have been measurable as well,
but its predicted location coincided by chance with that of the pen
mark (i.e.~``The Blob'') on the HIRES Tektronic's CCD.
The weaker {\FeII} transitions were covered by the spectrum, but were
not found to the 5$\sigma$ significance level.  
Their limits are consistent with the {\FeII} $\lambda 2383$ detection.

In the lower panel of Table~\ref{tab:systems}, the measured
rest--frame equivalent widths, column densities and Doppler $b$
parameters are presented.  
The measured {\MgII} doublet ratio is $2.0\pm0.1$, though it may 
be closer to 1.8, as we discuss below.
The {\MgII} and the {\FeII} lines are unresolved.
For unresolved lines, it is difficult to accurately determine the
column densities and $b$ parameters from profile fitting. 

We have performed extensive VP fitting simulations of the {\MgII}
doublet for this system.
The constraints for adopting the best model were the measured
{\MgII} $\lambda 2796$ equivalent width and the doublet ratio.
Using the curve of growth, we explored a grid of column densities and
$b$ parameters that were consistent with the measured equivalent width
of the {\MgII} $\lambda 2796$ transition.
The grid range was $12.1 \leq \log N({\MgII}) \leq 14.2$~{\cm2},
corresponding to $4.0 \geq b({\MgII}) \geq 0.4$~{\kms}.
The increments in column density were 0.1 dex.
For each grid location, 500 spectra were simulated, convolved with the
HIRES instrument spread function, sampled at the HIRES pixelization,
and degraded to the signal--to--noise ratio of the observed data.
The simulation output consisted of the {\MgII} $\lambda 2796$
equivalent width, the doublet ratio, and the VP column densities and
$b$ parameters from {\minfit}.

For $b\leq1$~{\kms}, we found that we could not recover the measured
equivalent width, nor the doublet ratio; they both decrease
dramatically with decreasing $b$.
This is due to the finite pixelization of HIRES.
As $b$ is reduced, the line depth increases.
As the pre--instrumentally broadened line width drops below that of a
single pixel, saturation losses dominate.
It could be argued that the measured equivalent width already reflects
this (that the absorption is actually stronger than the
measured value) and that larger equivalent widths should be explored.
Fortunately, the VP fits recovered the input $b$ values to both high
accuracy and precision over the full range explored.
For $b\leq1$~{\kms}, no matter the value of the equivalent width, the
doublet ratio could never be made consistent with the data (within
3$\sigma$).
For the lower limit on $b$, we adopted the criterion that the measured
doublet ratio would be a 3$\sigma$ outlier in the distribution of
simulated VP fits.
For the upper limit on $b$, we adopted the criterion that the measured
$b$ would be a $3\sigma$ outlier from the mode of the fitted $b$
distribution.

A caveat is worth noting.
We also explored simulations in which the {\MgII} doublets were fit
individually, rather than simultaneously.
From this we concluded that the observed $\lambda 2803$ transition is
likely compromised by a possible flat fielding artifact in its blue
wing.
This is consistent with the visual appearance of the data in
comparison to the many simulated spectra and with the measured $\chi
^{2}_{\nu}$ for the VP fits.
The value of $\chi ^{2}_{\nu}$ was 1.29 with $\nu = 74$ when all three
transitions were fit simultaneously.
If just the {\MgII} doublet was fit, then $\chi ^{2}_{\nu} = 1.47$
with $\nu = 51$.
This value is dominated by the ``poorer'' fit to the $\lambda 2803$
transition, which by itself was  $\chi ^{2}_{\nu} = 2.13$ with
$\nu = 24$.
In contrast, the VP fit to just the $\lambda 2796$ transition yielded
$\chi ^{2}_{\nu} = 0.97$ with $\nu = 24$.
The effect of this residual flux in the $\lambda 2803$ transition was
to push the measured $b$ value down to 1.5~{\kms}, when the doublet
was fit simultaneously.
When the observed $\lambda 2803$ transition was omitted from the VP
fit to the data, the resulting {\MgII} $b$ parameter was consistent
with the mean of the simulations.
Based upon these considerations, we have omitted the observed $\lambda
2803$ transition from the VP fit results; the measured $b$ parameter
used for interpreting the simulations was obtained by a fit to the
$\lambda 2796$ transition only.
Since we have adopted the assumption that the observed $\lambda 2803$ 
transition has been compromised, we have adopted the ``best'' doublet 
ratio from the simulations, ${\DR}=1.8\pm0.1$.
This implies that the equivalent width of the $\lambda 2803$ line is 
slightly larger than that formally measured from the data.
If the {\MgII} {\DR} is 1.8, then rest frame $\lambda 2803$ equivalent
width is $\sim 0.023$~{\AA}.

The adopted VP column densities are $\log N({\MgII}) =
12.24\pm0.09$~{\cm2} and $\log N({\FeII}) = 12.29\pm0.08$~{\cm2}.
Their respective $b$ parameters $b_{\rm tot}({\MgII}) =
2.2\pm0.5$~{\kms} and $b_{\rm tot}({\FeII}) = 2.3\pm1.6$~{\kms}. 
As with the $z=0.6428$, the value of $b_{\rm turb}/b_{\rm tot}$ for
the system could be determined from the ratio of the atomic masses of
iron and magnesium (see eq.~[\ref{eq:btot}] and
eq.~[\ref{eq:bXofbtot}]), but the uncertainties are too large to
directly place useful limits on $b_{\rm turb}$.

Because of the low resolution of the FOS spectrum, we could not obtain
estimates of the {\HI} column density and $b$ parameter directly from
the data.  
However, the wavelength range over which the Lyman limit break could
be observed is present in the spectrum at 1760.9~{\AA}.
There is no apparent flux decrement at the expected position of the
break. 
However, the signal--to--noise ratio is low, $\sim 5$, and this places
a $3\sigma$ limit of 1.6 on the flux ratio across the break.
This corresponds to an upper limit $\log N({\HI}) \sim 16.5$~{\cm2}.
The {\Lyb} transition is covered at 1981.2~{\AA}, but the region is
dominated by the $z=1.1537$ {\Lynine} line at 1983.1~{\AA}, so {\Lyb}
does not provide a constraint on $N({\HI})$.
If the {\HI} and the {\MgII} arise in the same physical locations in
the absorber, then the total {\HI} $b$ parameter is constrained to be
$1.8 \leq b({\HI}) \leq 13.0$~{\kms} (see Figure~\ref{fig:z09315}a),
including the spread introduced by the uncertainty in the measured
$b_{\rm tot}({\MgII})$.
The lower limit corresponds to the case in which turbulence or bulk
motions dominate the line broadening and the upper limit corresponds
to a thermal motions scaling, $\sqrt{24} b_{\rm tot}({\MgII})$.
From the curve of growth, we have estimated that the inferred $b$
range translates to a neutral hydrogen column density range of 
$13.6 \leq \log N({\HI}) \leq 16.5$~{\cm2}, where we have adopted the
upper limit from the Lyman limit break constraint.
We have included the spread introduced by the uncertainty in the
measured $W_{\rm r}({\Lya})$ in this estimate,  which dominates over
the uncertainty in $b_{\rm tot}({\MgII})$.

As with the $z=0.6428$ absorber, we have systematically searched the FOS
spectrum for other transitions associated with the $z=0.9315$ absorber
using the detection technique described by Schneider \etal (1993).
Details of the search are presented in Appendix~\ref{app:fossearch}
and selected results are tabulated in Table~\ref{tab:limits}.

%%%%%%%%%%%%%%%%%%%%%%%%%%% SECTION %%%%%%%%%%%%%%%%%%%%%%%%%%%%%%%%%%%%%%%%%%%

\section{Modeling The Absorbers}
\label{sec:models}

In order to better understand the two absorbers, we have attempted
to constrain their physical conditions, i.e.~ionization and chemical
conditions, non--thermal motions, and sizes.
In particular, we are interested in the relationship between the
ionizing flux, whether it is UVB or stellar/galaxy, and the inferred
metallicity/abundance pattern.
Taken together, constraints on these two quantities may reveal a great
deal about the origin, history, and local environment of the
absorbers.

We have modeled the absorbers as single--phase photoionized clouds
using CLOUDY (\cite{ferland}).  
The clouds were assumed to have constant density and plane--parallel
geometry. 
A grid of models were produced; for each model cloud the specified
physical conditions were (1) the ionizing continuum shape and
intensity, (2) the abundance pattern of the metals, and (3) the cloud
neutral column density, $N({\HI})$.
These input quantities constitute the biggest uncertainties in
modeling the absorbers.
We used CLOUDY in optimize mode, in which the residuals between the
model and the measured {\MgII} and {\FeII} column densities were
minimized.
The two quantities allowed to vary (optimized) were (1) the
metallicity of the assumed metal abundance patterns, and (2)
the total hydrogen density, $n_{\rm H}$.
For those ionization species for which column density upper limits 
were available, we applied the upper limits to the models.

\subsection{The Photoionizing Sources}

The two physical conditions within the absorbers that are the most
telling of its formation history are their abundance pattern/metallicity
and their photoionization source, either local stellar radiation or
the UV background (UVB).
In fact, the inferred chemical conditions are sensitive to the
intensity and shape of the ionizing flux continuum.
There are several scenarios and we address a few of the more obvious
ones below.

\subsubsection{The UVB Scenario}

The two absorbers, whether associated with galaxies or not, may have
photoionization conditions dominated by the UVB. 
To model this possibility, we have employed the UVB spectrum of Haardt
\& Madau (1996\nocite{handm96}), where the intensity has been
normalized at $z=0.5$ and $z=1.0$ for the $z=0.6428$ and the
$z=0.9315$ absorbers, respectively.
The Haardt \& Madau UVB spectrum accounts not only for the UV flux
emitted by QSOs and active galactic nuclei, but also for additional UV
flux due to the reprocessing of soft X--rays (also from the QSOs and
active galactic nuclei) in intervening absorbers at all redshifts.

The UVB spectra are shown in Figure~\ref{fig:uvb}, where only a
select range of energies is shown.
Also illustrated are the locations of the ionization potentials of a
few key ionization species.
The relevant ionization potentials that we are studying are all just 
above 1~Ryd, the ionization potential of {\HI}.
{\MgII} and {\FeII} have ionization potentials of 1.11 and 1.19~Ryd,
respectively.
The {\CII} ionization potential is 1.79~Ryd, and for {\CIII} is
3.52~Ryd. 
We mention {\CII} and {\CIII} because they probe the {\HeI} edge, at
1.81~Ryd, and because we have limits on the {\CII} and {\CIII}
absorption strengths.

\subsubsection{The Stellar/Galaxy Scenario}
\label{subsec:stars}

These particular clouds could be embedded within galaxies that are
aligned with the QSO on the plane of the sky (zero impact parameter),
or they could be in the outskirts of the galaxies, i.e.~in the
extended halo or outer disks.
In the latter scenario, the radiating stars can be treated as if they
are all equidistant from the clouds.
For any stellar/galaxy scenario, the number of stars and their
spectral types, metallicities, and distances (quantities that
determine the intensity and continuum shape of the ionizing flux) 
must be consistent with known objects in the universe.

The stellar/galactic UV flux could arise from a late--type
solar--metallicity stellar population, which would give rise to
a rapidly falling continuum with large {\HI} and {\HeI} breaks.
A ``soft'' spectrum is required by the observed upper limits on the high
ionization species. 
A central question defining the scenario is: what level
could stellar/galaxy flux be contributing to the UVB or
be completely dominating the UVB?
We have explored this question and have outlined the astrophysical
principles in Appendix~\ref{app:numstars}.
We constructed stellar/galactic CLOUDY grids that included three
galactic spectral energy distribution models over a range of
intensities and covered cases in which the stellar/galaxy flux was
progressively stronger compared to the UVB and the case in which the
UVB was locally ``extinct''.

For the ``dominant stellar--type'' scenarios, in which the cloud could
be near a dominant single star, we used Atlas stellar models
(\cite{kurucz}). 
We produced a grid of optimized CLOUDY models using solar metallicity
stars with $T_{\rm eff} = 6,000$, 10,000, 15,000, 20,000, 30,000~K,
and $\log g = 4.4$ (the solar value).
The spectral shape is not sensitive to the surface gravity, but
is quite sensitive to the metallicity and the effective surface
temperature.
The continuum falls more rapidly toward the UV for solar metallicity
stars, so these stars have ``softer'' continua than what would be
expected in a low metallicity early--type galaxy.

To account for various metallicities and/or stellar populations, we
also produced optimized CLOUDY grids using synthetic galaxy spectra.
We used a 12~Gyr single--burst Worthey (1994) model with metallicity
$[Z/Z_{\odot}]=-0.7$, and a somewhat younger Worthey model with a
8~Gyr single burst with metallicity $[Z/Z_{\odot}]=-2$.
We also used a later--type galaxy model from Bruzual \& Charlot
(1993\nocite{bruzual}) with an exponentially decreasing star formation
rate (SFR).
This model is a 16~Gyr stellar population with 1\% of the total
star--forming mass in stars after a Gyr.
These models serve to bracket a reasonable spread in galaxy spectral
properties, given that an extreme scenario such as a star bursting
galaxy can be ruled out for two reasons. 
First, a star burst spectrum would highly ionize the gas, giving rise
to {\SiIV} and {\CIV} absorption out to a galactocentric distance
of $\sim 100$~kpc (cf.~eq.~[2] of \cite{giroux}).
In fact, we found that it was difficult to not produce too much
{\SiIV} and {\CIV} even with the exponential SFR model.
Second, images of the {\qso} field, in which point spread function
removal of the QSO has been performed to high accuracy, reveal no
unidentified luminous objects with the characteristics of a star
bursting galaxy to a limiting $K$ magnitude of $\sim 20.5$.

\subsection{The Metallicity and Abundance Pattern}

Different chemical enrichment histories and different environments can
give rise to a wide variety of chemical and ionization conditions.
Given the possible high iron to magnesium abundance ratio in these
absorbers, it is reasonable to assume that the clouds could arise in
or near galaxies.  
Thus, to better understand the origin of the absorbers, we modeled
three abundance patterns that are taken from typical gaseous objects
in galaxies.
The first is the solar abundance pattern, taken from Grevesse \&
Anders (1989\nocite{grevanders89}) and Grevesse \& Noels
(1993\nocite{grevnoels93}).

In the interstellar medium (ISM), both magnesium and iron deplete onto
dust grains (cf.~{\cite{lauroesch}; \cite{savagearaa}).
Since the main constraints on the CLOUDY optimization are the measured
{\MgII} and {\FeII} column densities, we also considered the effects of
heating and cooling by grains and the dust depleted abundance patterns
of two common interstellar environments.
We used the {\HII} abundance depletion pattern taken from Baldwin
\etal (1991\nocite{baldwin91}), Rubin \etal
(1991\nocite{rubin91}), and Osterbrock, Tran, \& Vielleux
(1992\nocite{osterbrock92}).  
For this abundance pattern we used the ``large--R'' grains
(\cite{baldwin91}), which are characterized by a more or less grey UV
extinction.
Like the solar abundances, the {\HII} pattern, which has
$[\hbox{Mg/Fe}] = -0.07$, provides a good template for an iron--group
enriched chemical evolution history.

We have also explored the possibility that these absorbers may
actually be $\alpha$--group enhanced.
Thus, we used the abundance patterns reported by Cowie \& Songaila
(1986\nocite{cowie86}) for the cold and warm phases of the ISM.
This pattern is characterized by $[\hbox{Mg/Fe}] = 1.23$, and overall
enhanced $\alpha$--group elements.  
The dust grains used in these ISM models are a mixture of the graphite
and silicates.

The presence of grains also effects the cooling and heating balance,
and thus the ionization balance, of the clouds.
As mentioned, the metallicity, or the scaling factor of the input
abundance pattern for elements heavier than helium, was allowed to
vary.
As the metallicity in a cloud is increased by factors of a few, the
cooling rates are dramatically increased and the cloud equilibrium
temperatures drop significantly.
For some clouds, the optimal metallicity was 10 to 100 times the
initial input and the cloud equilibrium temperature in the
100~K range.

\subsection{Turbulent and/or Bulk Motions}

It is not possible to obtain a direct determination of the {\HI}
column densities for the two {\MgII} absorbers.
For each, the extreme range of plausible {\HI} column densities can be
estimated from the measured {\MgII} $b$ parameter for the assumption
of a thermal line broadening [lower $N({\HI})$ limit] or turbulent
broadening [upper $N({\HI})$ limit].
The range turns out to be large, $13.8 \leq \log N({\HI}) <
17.0$~{\cm2}, when the uncertainties in $b({\HI})$ and $W_{\rm r}
({\Lya})$ are considered.
We have modeled clouds with $\log N({\HI}) = 14.50$, 15.50, 15.75,
16.00, 16.25, 16.50, 16.75, 17.00, and 17.50~{\cm2}.
As will be shown, there were no $\log N({\HI}) \leq 15.75$~{\cm2} cloud
models consistent with the data.
Since, for a known equivalent width, the curve of growth provides a
direct relationship between the {\HI} column density and $b$
parameter, both quantities should be considered for constraining viable
cloud models.
Thus, we have developed a technique to further constrain the
acceptable model cloud properties using the relationship between the
turbulent and thermal line broadening mechanisms.

Using the equilibrium temperature of the CLOUDY models and the
$b_{\rm tot}$ of {\MgII}, one can narrow parameter space to those
models that are self--consistent in $b_{\rm turb}/b_{\rm tot}$,
temperature, and $N({\HI})$.
Here, $b_{\rm turb}$ is the non--thermal contribution to the line
broadening.
Assuming the contributions are Gaussian, $b_{\rm tot}$ is written,
\begin{equation}
b_{\rm tot}^{2} = b_{\rm therm}^{2} + b_{\rm turb}^{2} 
                = \frac{2kT}{m} + b_{\rm turb}^{2} ,
\label{eq:btot}
\end{equation}
where $T$ is the equilibrium kinetic temperature, and $m$ is the ion
mass.
In reality, non--thermal motions are not expected to be Gaussian
(see discussion in \S\ref{sec:caveats}), so eq.~[\ref{eq:btot}] is
simply a parameterization of the relative contributions.
Defining $f=b_{\rm turb}/b_{\rm tot}$, the kinetic temperature of the
cloud is written,
\begin{equation}
T \simeq 1450 \left( 1 - f^{2}\right) b^{2}_{\rm tot}({\MgII}) \quad
\hbox{K} .
\label{eq:tofbtot}
\end{equation}

In Figure~\ref{fig:temp}, the predicted kinetic temperature for
absorbers are shown as a function of $f$.
Schematically superimposed are the kinetic temperature ranges of
typical gaseous structures in the Galaxy (\cite{savagearaa}; 
\cite{fitzpatrickII}; \cite{deoagn2};  \cite{spitzer}).
Also shown is the inferred range of {\MgII} absorbers, as observed
using QSO absorption lines (\cite{mythesis}; Churchill \etal
1998\nocite{cvc98}).
This diagram is discussed further in \S\ref{sec:results}.

The $b$ parameter of each species is required in order to obtain
its estimated column density from the curve of growth.
Usage of the curve of growth method is consistent with our assumption
of single--phase clouds, given that the velocity structure of the
low ionization absorption profiles are not complicated and are
suggestive of a single absorbing component.
Assuming that all species are subject to the same thermal and
non--thermal conditions, the total Doppler $b$ parameter of any
species, X, can be obtained from,
\begin{equation}
b_{\rm tot}({\rm X}) = \left\{ \frac{A_{\rm Mg}}{A_{\rm X}}
  \left( 1 - f^{2} \right) + f^{2} \right\} ^{1/2} b_{\rm
  tot}({\MgII})  \quad \hbox{\kms} ,
\label{eq:bXofbtot}
\end{equation}
where $A$ is the nucleon number ($A_{\rm Mg} = 24$).

\subsection{Application of Constraints to Models}

In the absence of direct measurements of $N({\HI})$, the most acceptable
cloud models are those that are self--consistent in that their
$N({\HI})$, temperatures, and $f$ are simultaneously consistent with
those allowed by the data.
With the above formalism in hand, the application for constraining a
given cloud model is as follows: 
The range of acceptable $f$ values parameterize the clouds, and
give the inferred non--thermal component to the line broadening.
The range is determined from the model cloud kinetic temperature, $T$,
using a modified version of eq.~[\ref{eq:tofbtot}], 
\begin{equation}
f_{(\mp)} = 1 - \frac{1}{b_{(\pm)}^{2}({\MgII})}
\left( \frac{T}{1450} \right) ,
\end{equation}
where $b_{(\pm)}({\MgII}) = b_{\rm tot}({\MgII}) \pm \sigma _{b}$.
The cloud temperature was taken as a simple average, given that the
model clouds are constant density by definition.
For clouds with $N({\HI}) \leq 16.5$~{\cm2}, the temperature was
constant as a function of depth.
For the higher column density clouds, ionization and temperature
structure was present, but by no more than 10\%.
Eq.~[\ref{eq:bXofbtot}] was then used to obtain the inferred $b_{\rm
tot}({\HI})$, as illustrated in Figure~\ref{fig:z06428}a and
Figure~\ref{fig:z09315}a.
Using curve of growth analysis, the range of acceptable
hydrogen column densities is obtained from the measured {\Lya}
equivalent width (and its uncertainty) and the range of acceptable
$b_{\rm tot}({\HI})$.

\subsection{Caveats Regarding the Model Design}
\label{sec:caveats}

The quoted upper limits on the ionization species covered by the FOS
spectrum were estimated assuming a single--phase isothermal cloud.
There are {\it a priori\/} reasons that we have adopted this
assumption.
First, if the absorbers are {\Lya} clouds that are pressure confined
by a warmer, more tenuous medium (more than one thermal phase giving
rise to absorption), they cannot be heated by the UVB
(\cite{guilbert}; \cite{donahue}).
On the other hand, under the assumption of photoionization by the UVB,
a very low ionization parameter, $U = n_{\gamma}/n_{\rm H} \leq
10^{-3.5}$, is required in order for the absorbing gas to give rise to
$N({\FeII}) \sim N({\MgII})$ (\cite{jbands}; \cite{mythesis}).
Therefore, for UVB photoionization, clouds giving rise to the {\MgII} and
comparable {\FeII} absorption must have low ionization conditions. 
If strong {\CIV} and {\SiIV} were to arise in a UVB photoionized
absorber exhibiting $N({\MgII}) \simeq N({\FeII})$, then the abundance
pattern would need to be very different from solar, with a $1-2$ dex
enhancement of iron or decrement of magnesium. 
Common chemical evolution paths do not give rise to this type of
abundance pattern.

Nonetheless, the possibility remains that the absorbers could be
multiphased in their ionization structures (perhaps they are not
pressure confined or are ionized by stellar/galactic flux), so that
the measured {\Lya} equivalent width may be due to a kinematic complex
of clouds in which only one is giving rise to the observed {\MgII} and
{\FeII} absorption.
As reported by SC96, {\Lya} clouds at $z=2.5$ are seen to exhibit
kinematic structure in {\CIV} absorption, with the highest velocity
components more highly ionized.
They suggest that this arises due to layered stratification in 
coalescing clouds.
If this were the case for the low redshift absorbers reported here,
then the upper limits provided by the FOS spectrum would not apply to
this higher ionization phase; the limits would be lower than those
quoted above on account of the expected larger $b$ parameters.

If a high ionization phase exists in either absorber, it has
apparently not been detected in the FOS spectrum, suggesting that
either its ionization conditions or abundance patterns are not giving
rise to strong absorption lines.
For the ionization conditions to give rise to a multiphase absorber,
the neutral hydrogen column density would need to approach the Lyman
limit value of $10^{17.3}$~{\cm2} to provide shielding to the low
ionization phase. 
If there is no surrounding hot phase associated with these absorbers,
it might be difficult to understand them in terms of pressure confined
clouds; they could either be gravitationally bound or not be in
dynamical equilibrium.
Observationally, some ambiguity still remains for the detection of
{\CIV} and {\SIV} in the $z=0.6428$ absorber.
This uncertainty could be checked by higher resolution and higher
signal--to--noise ratio observations with STIS/HST.

A main point here is that the physical region of the absorbers
giving rise to the {\MgII} and {\FeII} must have insignificant {\CIV}
and {\SiIV} abundances.
A second main point is that the estimated {\HI} column density in this
region would be an upper limit to this low ionization phase, since
some fraction of {\HI} would be arising in the high ionization phase
as well.
Unless the path length through a possible high ionization phase is far
greater than that of the low ionization phase, the estimated ranges of
the {\HI} column densities for the two absorbers cannot be
significantly incorrect.

There is also the possibility that the spatial alignment of the
neutral hydrogen in these absorbers may not directly coincide with
that of the ionized magnesium (ionization structure), since the
ionization potential of {\MgII} and {\HI} are slightly different
(\cite{pbprivcomm}). 
Again, this would imply that the {\HI} column densities in the region
where {\MgII} and {\FeII} are giving rise to absorption is smaller
than the estimated values.

The presence of dust can effect the modeling in two ways.
First, for the model clouds using the {\HII}/ISM dust--depleted
abundance pattern, there is a breakdown in the inferred metallicity
when the temperatures are below $T\sim 1000$~K.
The condensation temperature (the temperature at which 50\% of an
element is removed from the gas phase due to depletion) of both
magnesium and iron is $\sim 1300$~K.
This means that these cool clouds would have a significantly higher 
depletion than the warmer clouds, and this has not been accounted for
in the CLOUDY modeling.
The implication is that the optimized metallicity is in fact
an underestimate of the gas--phase abundances, given that the
depletion would be larger than that input into the model.\footnote{These
$T\leq 1000$~K model clouds are already characterized by quite large
metallicities, even without accounting for increased condensation.}
Second, if dust is present, it can seriously modify the intensity and
shape of the incident UV spectrum, which would have implications for
the inferred source of the radiation from the photoionization models.
We discuss this point further in \S\ref{sec:stellarfail}.

Central to the model interpretation is the assumption that the line
broadening is governed by Gaussian distribution functions for the
particle velocities.
N--body simulations with hydrodynamics (\cite{norman}; \cite{rad};
\cite{zhang}) have shown that the absorbers are filamentary structures
and that the concept of an absorbing ``cloud'' is not altogether valid.
Often, the simulations result in gas which is collapsing towards
(or expanding along) a filament; this results in hydrodynamic features
and shocks in which the distribution function of velocities is not
Gaussian.
If the {\MgII} absorption profiles give any indication of the
velocity distribution function, then the absorbers are not
inconsistent with a Gaussian; in fact, very quiescent gas is suggested
by the HIRES profiles.

A final caveat is that we did not include turbulence physics in the
CLOUDY models (only micro--turbulence and not bulk motions could have
been modeled using CLOUDY).  
Thus, turbulence is not treated self consistently within the framework
used to infer the absorber physical conditions.
The inclusion of turbulence would result in an additional pressure
source, which would effect the balance between the model cloud density
and its depth.
However, the model cloud densities were allowed to vary in an
optimized fashion, and since the cloud depth is adjusted with each
equilibrium calculation the pressure adjustment has little effect.
The second consequence of modeling turbulence is that the line--center
optical depths of absorption transitions decrease, while they
increase in the line wings.
However, since the majority of the model clouds in our grids were
optically thin to neutral hydrogen, any change in the line profile
shape has little effect on the model cloud equilibrium.

%%%%%%%%%%%%%%%%%%%%%%%%%%% SECTION %%%%%%%%%%%%%%%%%%%%%%%%%%%%%%%%%%%%%%%%%%%

\section{Model Results}
\label{sec:results}

In the final analysis, only the scenario in which the model clouds are
photoionized by the UVB was both astrophysically plausible and
consistent with the data.
Here, we focus on the results for the UVB scenario with the solar
and {\HII}/ISM abundance patterns and then address to what level the 
stellar/galactic scenarios can be ruled out as viable ionizing
sources.
We defer discussion of the implications of the model results until
\S\ref{sec:discuss}.

The optimized metallicities and densities are given for
both abundance patterns in Tables~\ref{tab:hm06428} and
\ref{tab:hm09314}.
Cloud models with the $\alpha$--group enhanced abundance pattern
(\cite{cowie86}) did not converge within the allowed uncertainties in
the {\MgII} and {\FeII} column densities.
Thus, we conclude that neither the $z=0.6428$ nor the $z=0.9315$
absorber have $\alpha$--group enhanced abundance patterns.
This implies a gas--phase $[\hbox{Fe/H}] \geq -1$ ({\cite{lauroesch};
\cite{savagearaa}), and possibly iron--group enrichment by Type Ia
SNe, though this remains somewhat controversial (Gibson \etal
1997\nocite{gibson}).
The $N({\HI})$ versus $f$ parameter space is illustrated in
Figures~\ref{fig:z06428} and \ref{fig:z09315} for the data and for the
model clouds.
The adopted ranges for the absorber {\HI} column densities are defined
by the overlap of the allowed ranges constrained by the data and by
the CLOUDY models.
Interestingly, the results indicate that the clouds have a substantial
non--thermal line broadening (i.e.~they are turbulent or are
undergoing differential bulk motions).
However, it is striking that the {\MgII} profiles reveal a 
velocity dispersion of only a few {\kms}.
This provides a counter example to the expected large $b$ parameters
if the gas was not dynamically settled (as found in hydrodynamic
simulations of the {\Lya} forest).
The very quiet nature of these absorbers indicate that our assumption
of a well defined temperature for a settled gas ``cloud'' is well
founded.
A possible, though unlikely, counter example would be if the absorbers
were streaming filaments seen perpendicular to their elongation and
streaming motion.

Shown in Figure~\ref{fig:temp} are the predicted kinetic temperatures
of the absorbers as a function of $b_{\rm turb}/b_{\rm tot}$.
Temperature and turbulence limits may provide clues to the nature of
the absorbing gas when the range of inferred properties are compared
to gaseous objects typically found in galaxies.
Planetary nebulae have $T \geq 40,000$~K (\cite{deoagn2};
\cite{spitzer}).
If the $z=0.6428$ absorber has $f \leq 0.4$, then its inferred
temperature is consistent with that of a planetary nebulae.  
Typical expansion velocities of these objects have line widths of
$\sim 20-30$~{\kms} (\cite{deoagn2}).
Since the {\MgII} and {\FeII} Doppler parameters would also reflect
the expansion velocities, it is highly unlikely that the $z=0.6428$
absorber arises in a planetary nebula.
For $0.4 \leq f \leq 0.9$, the $z=0.6428$ absorber temperature is
consistent with that of individual clouds in complex {\MgII}
systems (\cite{mythesis}; Churchill \etal 1998\nocite{cvc98}). 
Both absorbers are consistent with the temperature range of {\HII}
regions and warm {\HI} clouds, $7000 \leq T({\HII}) \leq 14,000$~K and
$4000 \leq T({\HI}) \leq 7000$~K, respectively (\cite{fitzpatrickII};
\cite{deoagn2}; \cite{spitzer}).
For the $z=0.9315$ absorber, the inferred $b_{\rm turb}/b_{\rm tot}$
would be $f \leq 0.8$, and for the $z=0.6428$ absorber $f$ would be
confined to the narrow range $0.90 \leq f \leq 0.95$.
For these $f$ values, the non--thermal broadening would be roughly 
$b _{\rm turb}({\MgII}) \leq 5$~{\kms} and $\leq 2$~{\kms} for the
$z=0.6428$ and the $z=0.9315$ absorber, respectively.
The typical sound speeds in {\HII} and {\HI} clouds are $\sim
10~{\kms}$ (\cite{spitzer}).
If these absorbers are {\HII} or {\HI} clouds similar to those found
in the Galaxy, the line of sight non--thermal broadening is well below
that expected for propagating disturbances in the clouds.
If the absorbers are dominated by turbulent motions ($f>0.98$), then
they must have $T \leq 150$~K.
Diffuse ISM clouds have typical temperatures in the range $30
\leq T \leq 150$~K (\cite{spitzer}).
In this regime, it would be more likely that the broadening was
dominated by bulk flows rather than internal turbulence, given that
the turbulent motion would propagate at the $\sim 0.1$~{\kms} sound
speed typical of diffuse clouds (\cite{spitzer}).

\subsection{The $z=0.6428$ Absorber Properties}
\label{sec:0.6428}

Assuming the solar abundance pattern, the neutral hydrogen of the
$z=0.6428$ model cloud is in the range $16.3 \leq \log N({\HI})
\leq 16.8$~{\cm2}.   
The range of $b_{\rm turb}/b_{\rm tot}$ is $0.85 \leq f \leq 0.93$,
which correspond to the kinetic temperatures $13,000 \geq T \geq
6500$~K.
The metallicity and model cloud density are $-0.2 \geq
[Z/Z_{\odot}] \geq -0.7$, and $0.01 \leq n_{\rm H} \leq
0.02$~cm$^{-3}$, for the range of $N({\HI})$.
For the {\HII} abundance pattern, the inferred neutral hydrogen column
density is slightly higher, in the range $16.7 \leq \log N({\HI})
\leq 17.2$~{\cm2}.   
The range of $b_{\rm turb}/b_{\rm tot}$ is $0.90 \leq f \leq 0.95$,
making this cloud kinetic temperature somewhat lower than the 
solar abundance model. 
The density and metallicity are $n_{\rm H} \simeq
0.008$~cm$^{-3}$ and $+0.4 \geq [Z/Z_{\HII}] \geq 0.0$, for the
range of $N({\HI})$.
It could be that this cloud has very enhanced metallicity with an 
{\HII} abundance pattern, but this is a far reaching suggestion given
the range allowed by the solar abundance pattern.
Still, the cloud is inferred to have gas--phase $[\hbox{Fe/H}] \geq
-1$.

These model clouds are difficult to understand in terms of objects
typical of the Galactic disk or the Magellanic Clouds.
For one, the typical $N({\HI})$ observed in Galactic objects is $\log
N \geq 19.5$~{\cm2} (\cite{savagearaa}), two or more orders of
magnitude greater than what is inferred for this absorber.
Second, the typical density of $T \sim 10,000$~K clouds (warm low
density medium) is $\left< n_{\rm H} \right> \sim 0.2$~cm~$^{-3}$
(\cite{spitzer}), which is higher than allowed by the optimal models.
Third, the inferred gas--phase abundances in the the warm and cool disk
are $[\hbox{Fe/H}] \sim -1.1 $ and $ \sim -2.1$, respectively
(\cite{savagearaa}).
These fall well below those predicted by the models.
However, the metallicity range of the solar abundance pattern model is
consistent with $[\hbox{Fe/H}] \sim -0.6$ found for the Galactic Halo.

\subsection{The $z=0.9315$ Absorber Properties}
\label{sec:0.9315}

The $z=0.9315$ model cloud appears to have a high gas--phase
metallicity, whether it be super solar or an enhanced {\HII} pattern.
Assuming the solar abundance pattern, the neutral hydrogen is in the
range $15.8 \leq \log N({\HI}) \leq 16.3$~{\cm2}.   
Note that this is consistent with the upper limit of 16.5~{\cm2}
inferred from the lack of a Lyman limit break in the FOS spectrum.
The range of $b_{\rm turb}/b_{\rm tot}$ is $0.62 \leq f \leq 0.94$,
which correspond to the kinetic temperatures $4600 \geq T \geq 870$~K.
The metallicity and model cloud density are $+0.7 \geq
[Z/Z_{\odot}] \geq +0.1$, and $0.2 \leq n_{\rm H} \leq
0.4$~cm$^{-3}$, for the range of $N({\HI})$.
The optimal solar abundance model cloud is relatively dense with
up to five times solar abundance.
The inferred $[\hbox{Fe/H}]$ is even greater for the {\HII} abundance
pattern model.
The neutral hydrogen column density is slightly higher, in the range
$16.0 \leq \log N({\HI}) \leq 16.5$~{\cm2} (also consistent with the
lack of a Lyman limit).
The range of $b_{\rm turb}/b_{\rm tot}$ is $0.75 \leq f \leq 0.97$,
which corresponds to the kinetic temperatures $3300 \geq T \geq 450$~K.
There is an inversion in the cooling curve at $T\sim 2000$~K, which
implies that this absorber cannot be stable across the full range of
allowed temperatures.  
It must either be a few thousand degrees or several hundred degrees.
The density and metallicity are $n_{\rm H} \simeq
0.1$~cm$^{-3}$ and $+1.6 \geq [Z/Z_{\HII}] \geq +0.7$, for the
respective $N({\HI})$.
This is a metallicity enhancement of five to 40 times over the typical
values seen in Galactic {\HII} regions (\cite{baldwin91};
\cite{rubin91}; Osterbrock \etal 1992\nocite{osterbrock92}).   
To date, no other intervening QSO absorption system with such a
high metallicity has been reported.

\subsection{Why the Stellar/Galaxy Scenarios Fail}
\label{sec:stellarfail}

In what follows, we discuss the difficulties with the stellar/galaxy
scenarios.
The constraints were a trade off between the number of stars, or their
number density, and the stellar population.
The former is constrained by astrophysics, assuming a non--extreme
stellar environment, and by the imaging data.
The latter is constrained by the absorption line data, which have
limited the UV ionizing flux to late--type stars and/or early--type
galaxies.

First, consider the case in which the stellar/galactic flux contributes
to the UVB intensity.
In order for a stellar/galactic contribution to modify the properties
of a UVB model cloud, the stellar/galactic flux must exceed the UVB at
$\geq 1$~Ryd, particularly in the regions $1 \leq h\nu \leq 1.2$~Ryd
(from the {\HI} edge up to and including the {\FeII} ionization
potential).
As outlined in Appendix~\ref{app:numstars}, if the stellar population
is dominated by A0III and A0V stars, this requires $\sim 10^{12}$
stars confined to a region of space $\sim 1$~kpc in radius. 
This implies a stellar number density $n_{\ast} \geq 500$
stars~pc$^{-3}$, which is five orders of magnitude greater than the
density of A0V stars in the solar neighborhood (\cite{allen}).
The required number density increases dramatically for later spectral
types.
If, on the other hand, the stars are dominated by early--type B0V
(B0I,III) stars, then $\sim 10,000$ (1000) stars would be required in
a volume of radius 1~kpc.
For the main sequence stars, this corresponds to a number density
about 100 times greater than that of the solar neighborhood
(\cite{allen}).  
Only O stars can provide the UV flux necessary to match the UVB at
1~Ryd and have a number density of stars consistent with that of a
typical galaxy environment.
However, early--type stars give rise to high ionization absorption
properties, especially {\CIII}, {\CIV}, and {\SiIV}, so that the model
cloud conditions are not consistent with the data.

The 12 Gyr $[Z/Z_{\odot}] = -0.7$ Worthey (1994\nocite{worthey}) model
is characterized by a steep continuum slope with $\Delta \log \nu
f_{\nu} \sim -5$ from 1 to 1.2~Ryd.
This continuum shape is a smooth continuation of the {\HI} edge, so
that this galaxy model had to have $\nu f_{\nu} \geq 10^{8}$ times
that of the UVB at 5500~{\AA} before affecting the model cloud
properties.
Based upon the arguments presented in Appendix~\ref{app:numstars}, for
the expected distribution of main sequence and giant stars in these
galaxies (\cite{worthey}), the stellar number densities would be
extreme, $N_{\ast} \geq 10^{2}$ stars~pc$^{-3}$, or $\sim 10^{12}$
stars within a kpc.
Even under these extreme conditions, the 12 Gyr Worthey models yielded
$[\hbox{Fe/H}] \geq -1$.  

The 8~Gyr Worthey model with metallicity $[Z/Z_{\odot}]=-2$ is
characterized by a continuum with a smaller {\HI} edge of $\Delta \log
\nu f_{\nu} \sim -2$ at 1~Ryd and a power law with $\sim \nu ^{-7}$
out to $\sim 1.5$~Ryd (this is steep but not nearly as steep as the 12
Gyr model).
The flux must be elevated to $\nu f_{\nu} \geq 10^{5}$ times that of
the UVB at 5500~{\AA} before affecting the model cloud properties.
As the flux, $\nu f_{\nu}(0.17)$, is increased above $\sim
10^{4}$~ergs~{\cm2}~s$^{-1}$, the ratio $N({\HI})/N({\HII})$ very
quickly changes from unity to $\sim 10^{-5}$; the model cloud
becomes highly ionized and the limits on {\CIII} and/or {\CIV} are
exceeded (depending upon the specific absorber). 
To adopt the idea that the 8 Gyr Worthey galaxy is contributing to the
ionization conditions, we would be forced to accept a very narrow
range of acceptable $\nu f_{\nu}(0.17)$ arising from the galaxy. 
This range implies the absorbers would be embedded in the galaxy
itself (zero impact parameter), with an extremely unrealistic number
of stars.
Such a scenario is ruled out.

The exponential SFR spectrum of Bruzual \& Charlot (1993\nocite{bruzual})
has a continuum shape at $1-3$~Ryd that is similar to the Haardt \&
Madau UVB spectrum.  
Thus, as the galactic spectrum is slowly increased over the UVB, the
model cloud properties adjust such that the ionization parameter is
held constant (the cloud density increases, but the metallicity and
temperature do not change).
At $\nu f_{\nu}(0.17) \sim 0.1$~ergs~{\cm2}~s$^{-1}$, there is a
slow decrease in the ratio $N({\HI})/N({\HII})$; the model cloud
becomes progressively highly ionized. 
The dominant ionization species of magnesium and iron are then
{\MgIII} and {\FeIII}, respectively, and the {\CIV} and {\SiIV} column
densities exceed the limits allowed by the data.
The exponential SFR galaxy of Bruzual and Charlot is not a viable
photoionizing source.
The galaxy flux makes little modification to the inferred cloud
properties (does not modify UVB only models) for a large range of
intensities; once it does modify the UVB models, the cloud
properties are inconsistent with the data.
 
Now, consider the case in which the stellar/galactic flux is dominant.
We examined this by excluding the Haardt \& Madau UVB from the incident
flux.
This would require environments that are more extreme than those
presented above, or that are shielded from the UVB but not from the
stellar flux.
Is it possible that dust extinction could be playing a role?
For a dust absorption/scattering cross section of
$\sigma(\hbox{1~Ryd}) \sim 10^{-22}$~{\cm2} (\cite{mathis77}), the
total hydrogen column density [$N({\HI})+N({\HII})+2N({\rm H}_{2})$]
required for $\tau _{\rm dust} \sim 3$ at 1 Ryd is $\log N \sim
22.5$~{\cm2}.
Given the upper limits on $N({\HI})$ for the absorbers, this implies
highly ionized gas with $N({\HI})/N_{\rm tot}({\rm H}) \geq 10^{-5}$.
In all our model clouds, the fraction of molecular hydrogen never
increased above $f({\rm H}_{2}) \leq 10^{-6}$.

In principle, one could imagine an environment in which the absorbing
gas is embedded in a late--type dwarf galaxy that itself is
enshrouded in dust.
This scenario provides the steep continuum above 1~Ryd while not
requiring the spectrum intensity to be elevated above the UVB.
For some models it was possible to achieve $\log N({\HII}) \simeq
22.5$~{\cm2}, but even with no competition from the UVB, these
shielded models required high $\nu f_{\nu}(0.17)$.
Again, as outlined in Appendix~\ref{app:numstars}, this implies
unrealistic stellar densities.
It is not simply a matter of the continuum slope, but also the
intensity that dictates the cloud ionization conditions.
Interestingly, these models produced very low metallicities, with
$[Z/Z_{\odot}] \sim -3$.
All of the above stellar/galaxy models represent our failed attempt to
locate a place in parameter space consistent with low metallicity
clouds.

%%%%%%%%%%%%%%%%%%%%%%%%%%% SECTION %%%%%%%%%%%%%%%%%%%%%%%%%%%%%%%%%%%%%%%%%%%

\section{Discussion}
\label{sec:discuss}

Based upon the success rate of finding two {\MgII} absorbers out of 28
{\Lya} absorbers along the {\qso} line of sight, we must conclude that
the two absorbers are unique in some manner with respect to the {\Lya}
forest at large.
We note, as mentioned in \S\ref{sec:systems}, that not a great deal
can be quantified about the range of metallicities in the {\Lya}
forest at $z\le 1$ based upon our upper limits on the {\MgII} column
densities.
However, if our assumption of $b\sim30$~{\kms} is not applicable, as
found for the two {\MgII} absorbers, then our quoted estimates in the
range of metallicities would be quite wrong.
In fact, given that the upper limits on the metallicity are
super--solar for the $W_{\rm r}({\Lya}) \leq 0.3$~{\AA} absorbers, and
that it is not expected that many of the absorbers are super--solar,
we are lead to suggest that either the $b$ parameters of the $W_{\rm
r}({\Lya}) \leq 0.3$~{\AA} absorbers are significantly smaller than
30~{\kms} or the objects do not have a Gaussian velocity
distribution (simple curve of growth techniques are not applicable).
It could be that the uniqueness of the two {\MgII} absorbers is that
they {\it are\/} dynamically settled and have a small velocity
dispersion.

Other possibilities for the uniqueness of the {\MgII} absorbers, are:

(1) {\it Their ionization conditions are different}:
If the chemical enrichment histories of these absorbers are typical of
{\Lya} forest clouds, then it might be inferred that their ionization
conditions are governed by a local UV flux rather than the UVB.
However, our models lead us to conclude that there is nothing
``special'' about the photoionization conditions of these clouds;
they are best described as being photoionized by the UVB.
There is no evidence to suggest that these clouds are collisionally
ionized.

(2) {\it Their chemical conditions are different}:
If the ionization source is no different than that ionizing the other
26 {\Lya} systems we searched, then we are led to infer somewhat
unique chemical enrichment histories for these two absorbers.
If these absorbers are IGM/{\Lya} clouds, in that they are not
associated with galaxies, then the low metallicity results of SC96 at
higher redshift do not support the notion that these two absorbers
have undergone typical IGM chemical enrichment.
In other words, these two absorbers are not consistent with the
picture in which the IGM was enriched at high redshift by a single
burst of Population III stars that left the {\Lya} forest
$\alpha$--group enhanced\footnote{Depending upon the Population III
IMF, some pockets of the IGM may have experienced slow metallicity
build up due to late type stars.  However, we find this scenario to be
no different in name than if the process took place in ``galaxies''.}
with $[\hbox{Fe/H}] \leq -2$.
Thus, these absorbers could constitute a metal--rich minority of the
IGM/{\Lya} population (less than $\sim 10$\%).
It then becomes a question of understanding what environments and
evolutionary histories give rise to high metal content in some {\Lya}
clouds.
It is already known that at least some fraction of the {\Lya} forest
at $z<1$ are associated with galaxies (\cite{lebrunlya};
\cite{lanzetta}; Bowen \etal 1996\nocite{bowen}).

(3) {\it Both their ionization and chemical conditions are different}:
Inferring the chemical content of ionized absorbers requires
ionization corrections that are uncertain.
In the case of photoionization, the inferred conditions of the clouds
are very sensitive to both the intensity and shape of the ionizing
continuum.
We have explored this interplay between the chemical content of the
clouds and the properties of the UV ionizing flux, and have concluded
with some certainty that the ionizing field for the two absorbers is
constrained to have slope and intensity consistent with the UVB.
This is tantamount to saying that it is only the chemical conditions
that are inferred to be unique in the two absorbers.
 
Overall, it is difficult to understand these absorbers in terms of the
classic picture of {\Lya} forest clouds.
One problem is that their inferred {\HI} column densities are higher
than ``typical'' forest clouds.
A second is their high $[\hbox{Fe/H}]$ and iron--group enhance
abundance pattern.
Since the metal abundance patterns of these absorbers are not
$\alpha$--group enhanced, it is implied that their environments 
have been influenced by Type~Ia SNe yields (\cite{lauroesch};
\cite{fxt}; however, see {Gibson \etal 1997\nocite{gibson}).
It could be that these absorbers are associated with galaxies.
However, based upon imaging and spectroscopic studies, there are no
candidate high surface brightness (HSB) objects in the {\qso} field.

The extended luminous objects identified in the WFPC2 images (Figure~4
of LBBD) and ground--based image (Figure~3 of Steidel \etal 1995),
have now had their redshifts spectroscopically measured using LRIS on
the Keck~I telescope.
None of them have redshifts that match the two {\MgII} absorbers
(\cite{chuckprivcomm}).
Object \#5 in LBBD has been confirmed by Steidel and
collaborators to be the strong {\MgII} absorbing galaxy at
$z=1.1536$. 
At impacts greater than $120h_{75}^{-1}$ there are three galaxies with
$0.605 \leq z \leq 0.610$ (not presented in published images).
Thus, to a limiting of magnitude of $K\sim 20.5$, the limit of the
Steidel \etal image, there are no luminous candidates within $\sim
30${\arcsec} of the QSO.

Based upon the residuals following the point--spread function
subtraction of the QSO in both the WFPC and the ground--based images,
there is no evidence for luminous objects directly in front of the QSO
(``zero--impact'' absorbers).
However, dwarf galaxies of roughly $\leq 0.01~L^{\ast}$ at zero impact
cannot be ruled out.
In general, it seems unlikely that two absorbing galaxies, separated
by such a large redshift interval, would be aligned with the QSO on
the sky.
According to the work of Bowen \etal (1997\nocite{bowenleoI}), dwarf
spheroid galaxies similar to Leo I are not massive enough to have
halos that can contribute significantly to the metal line absorption
cross section of QSO absorbers seen at high redshift.
But the situation is not so clear overall, given the recent discovery
by BBLD of the saturated {\MgII} doublet associated with the
$z=0.072$ dwarf galaxy at impact parameter $2.7h_{75}^{-1}$~kpc.
The emission line properties of this dwarf (\cite{ccsdwarf}) suggest
that star formation in these objects may directly govern their gas cross
section.
If so, perhaps active star forming dwarf galaxies could contribute to
the overall metal line absorption cross section (cf.~\cite{york};
\cite{yanny}).
Naively, one would then expect that the abundance pattern arising from
a bursting dwarf would be $\alpha$--group enhanced.
Also, it is likely that UV ionizing flux from the newly formed O and B
stars would contribute to the ionization conditions in the absorbers,
which is not what we find.

If we are to assume that these two absorbers are associated with
galaxies of some type, and if we accept the lack of evidence for 
HSB candidates in the {\qso} field, then we
must explore the idea that low surface brightness (LSB) galaxies
(cf.~Bothun \etal 1997\nocite{bothun}) could be giving rise to the
absorbing gas.
Particularly, we are lead to consider the class of galaxies called 
``giant LSB galaxies'' (\cite{sprayberry93}; \cite{sprayberry95}).

Great progress in our knowledge of LSB galaxies, their number density,
sizes, metallicities, and luminosity function in the local universe has
been made over the past few years (\cite{deblok}; \cite{quillen};
\cite{dalcanton}; \cite{sprayberry97}; \cite{dejong};
\cite{sprayberry95}; \cite{mcgaugh}).
Dalcanton \etal (1997) find that LSB galaxies have a space density of
at least $n=0.03$~galaxies~$h_{75}^3$~Mpc$^{-3}$ and outnumber
comparable HSB galaxies by factors of $\sim 2$ or more.
LSB galaxies are a non--negligible component of the local universe
baryonic mass.
Sprayberry \etal (1995) find that LSB giants have larger disk scale
lengths than HSB galaxies of comparable total luminosity.
In the cases of F568--6 and UGC~6614, the luminous spiral arms 
extend to $80h^{-1}_{75}$ and $50h^{-1}_{75}$~kpc, respectively
(\cite{quillen}). 
The extent of the {\HI} disks for the general population of LSB
galaxies is seen to be roughly 2.5 times that of their $D_{25}$, the
diameters of their $\mu _{B} = 25$~mag~arcsec$^{-2}$ isophotes
(\cite{vanderhulst}).
If this scaling holds for F568--6, then it may have an {\HI} disk of
$\sim 200h^{-1}$~kpc.
In the case of the LSB galaxy $1226+0105$, the {\HI} disk may extend  
more than four times its $D_{25}$ (\cite{sprayberry95}).

Whatever structures give rise to these two absorbers, the {\HI} gas
must have a velocity dispersion of no more than $\sim 30$~{\kms}, as
dictated by the inferred upper limit on the $b$ parameter of the
$z=0.6428$ cloud.
The constraint is even as low as $\sim 15$~{\kms} based upon the
$z=0.9315$ cloud.
The best values of the cloud $b$ parameters are $\sim 12$~{\kms} and 
$\sim 9$~{\kms}, respectively.
These $b$ parameters fall {\it below\/} the lower cut offs in the
overall {\Lya} forest $b$ distribution (\cite{kim}; \cite{lulya};
\cite{hulya}).
Again, this suggest that these absorber possibly arise in a minority
sub--class of the overall {\Lya} cloud population.
In their study of the giant {\HI} disks of F568--6 and UGC~6614,
Quillen and Pickering (1997a) found that the {\HI} showed small
velocity dispersions of 10--30~{\kms} and 10--20~{\kms}, respectively,
as compared to 60--90~{\kms} measured for local HSB spirals
(\cite{vogel}; \cite{canzian93}).
These dispersions were measured among the spiral arms; it could be
that the extended outer disks are even more quiescent.
Even so, these values are consistent with the allowed ranges of the
two absorbers.
If a sub--population of the {\Lya} forest is arising in LSB galaxies,
they may be characterized by having $b$ parameters scattered about the
low end of the distribution.

The metallicities of several LSB galaxies have been measured using
{\HII} regions.
LSB galaxies with relatively smaller disk scale lengths are found to
have $[Z/Z_{\odot}] \leq -0.3$, and are therefore metal poor
(\cite{mcgaugh}).
However, McGaugh found near--solar and super--solar metallicities
for UGC~5709 and F568--6, respectively.
These two galaxies have large disk scale lengths, and classify as
giant LSB galaxies.\footnote{The giant, or large scale length, LSB
galaxies are defined by Sprayberry \etal (1995) to have $\mu _{B}(0) +
5 \log \alpha ^{-1} > 27$, where $\mu _{B}(0)$ is the central surface
brightness in the $B$ band and $\alpha^{-1}$ is the scale length in
$h^{-1}$kpc. The high metallicity LSB galaxies are seen to have
$\alpha^{-1} \geq 13h^{-1}_{75}$~kpc, which constitute $\sim 1/3$ of
the known giant LSB galaxies defined in Sprayberry et~al.}
Pickering and Impey (1995; \cite{impeyprivcomm}) have also found other
giant LSB galaxies have metallicities that scatter around solar.
These giant galaxies are found to have stellar surface densities at
least on the same order as their gas densities, which leads Pickering
and Impey to suggest that these galaxies have been forming stars
slowly.

We find these facts to be quite interesting in light of the two high
metallicity {\MgII} systems we have found.
It is well established that {\MgII} absorption with $W_{\rm r} \geq
0.3$~{\AA} selects the population of HSB galaxies
(\cite{csv96}; \cite{steideleso}; \cite{sdp94}; \cite{bb91}).
These galaxies appear to be ``normal'' in their morphologies and to
have luminosities greater than $\sim 0.05~L_{K}^{\ast}$.
By comparison, the total luminosities of giant LSB galaxies scatter about
$L^{\ast}$ (cf.~\cite{sprayberry95}).
Though LSB galaxies have low luminosity densities, their disks are
proportionally larger, giving them total luminosities on par with HSB
galaxies.
Thus, it seems reasonable that LSB galaxies could be part of a more 
general {\MgII} absorption selected galaxy population, where the
LSB galaxies are selected by the smaller {\MgII} equivalent widths.

In a recent survey to a limiting rest--frame equivalent width of
0.02~{\AA}, Churchill \etal (1997\nocite{crcv97}) found that {\MgII}
absorbers with $W_{\rm r}(\lambda 2796) \leq 0.3$~{\AA}  (hereafter
called ``weak {\MgII} absorbers'') account for $\sim 65$\% of all
{\MgII} absorbers (also see \cite{iapconf})
{\it Nothing is yet known about the type of luminous object they select};
none have luminous candidates to $\leq 0.06~L^{*}_{K}$ [assuming the
Freeman (1970\nocite{freeman}) surface brightness] in the survey of
{\MgII} absorbers by Steidel \etal (\cite{chuckprivcomm}). 
These systems also exhibit {\FeII} absorption; for the sample, $\left<
\log N({\FeII})/N({\MgII}) \right> = -0.3\pm 0.4$ (i.e.~they may
have $[\hbox{Fe/H}] \geq -1$).
Without information on their {\HI} absorption, we can only speculate
that some fraction of the weak {\MgII} systems have ionization and
chemical conditions similar to the two absorbers studied in this work.

If weak {\MgII} absorbers are selecting out a ``missing'' part of the
{\MgII} absorption selected galaxy population, LSB galaxies are a
logical candidate for this missing portion, particularly the class of
giant LSB galaxies, or ``Malin--cousins'', as designated by Sprayberry
\etal (1993, 1995).
These galaxies are disk galaxies, and these disks are observed to have
a lower neutral {\HI} surface density than HSB galaxies
(Bothun \etal 1997\nocite{bothun}).
As a result, the giant LSB galaxy disks have relatively quiescent
stellar evolution.
In fact, the general population of LSB galaxies show a trend of
increasing red color with increasing disk scale length
(\cite{sprayberry95}).
Quillen and Pickering (1997b) reported extremely red colors ($R-H = 2.2$
and $B-H=3.5-4.2$) for the two giant LSB galaxies UGC~6614 and 
F568--6, which suggest that they have a dominating old component in their
stellar populations.
These galaxies provide the precise type of environment in which there
has been ample time for iron--group enhancement and metallicity build
up in the gas phase of their disks.
Further, the quiescent nature of their disks leads us to suggest that
we should {\it not\/} expect to see the complex velocity structures seen
in the majority of the stronger {\MgII} absorption profiles
(\cite{mythesis}; Churchill \etal 1998\nocite{cvc98}).
Indeed, the low {\HI} surface density of these galaxy disks should
result in weaker {\MgII} absorption because there is less of the 
neutral hydrogen shielding required for {\MgII} to survive.
Also, the quiescent nature of the interaction between the gas and
stars in these galaxies suggests that the gas is not being stirred up,
which would generate erratic gas kinematics.
It may be that such processes facilitate the generation of a high
ionization layer around the disk, as seen in the Galaxy
(cf.~\cite{savagearaa}).
The small $b$ parameters inferred for the {\HI} and the apparent lack
of {\CIV} absorption in the two weak absorbers are consistent with a
quiescent disk.

If weak {\MgII} absorbers are selecting giant LSB galaxies, then
these absorbers provide a potential probe of the number density of
these massive galaxies.
LSB galaxy disks may grow from isolated 1--2$\sigma$ peaks in the
initial density fluctuation spectrum and may trace low density
extended dark matter halos in a relatively unbiased way
(Bothun \etal 1997\nocite{bothun}).
They also would provide a powerful probe of the chemical enrichment
history of LSB galaxies, which appear to evolve at a significantly
slower rate and may produce stars via conventional pathways [such as 
not within molecular clouds (cf.~Bothun \etal 1997\nocite{bothun})].

To date, there is not enough known about the number density and
gaseous cross sections of the class of giant LSB galaxies to compare
their $dN/dz$ directly to that of the weak {\MgII} systems, or to
place meaningful limits on their number evolution if we assume they
are selected by weak {\MgII} absorption (Bothun \etal
1997\nocite{bothun}; \cite{impeyprivcomm}).
Roughly, the overall population of LSB galaxies appear to follow a
trend such that those with larger disk scale lengths are observed to
have smaller central surface brightness (\cite{sprayberry95}).
Following this relation, there is a significant gap between Malin~1
and the remaining sub--population of giant LSB galaxies.
Malin~1 has a disk scale length a factor of 30 times greater than that
of  F568--6 and a central surface brightness a factor of 100 less than
F568--6.
Does the gap in this parameter space reflect a true break, suggesting
that Malin~1 is a rare galaxy type?
Or, is the gap an artifact of selection effects?
As Sprayberry \etal point out, it is important to explore this
parameter space in order to determine the size and number density
distributions of giant LSB galaxies.
If LSB galaxies are considered to be a natural extension of the HSB
galaxy luminosity function, and their disk scale lengths and central
surface brightnesses exhibit similar behaviors to those found for HSB
galaxies, then the region of central surface brightness -- scale
length parameter space giving rise to Malin--type LSB galaxies is 
continuously populated (\cite{slinderprivcomm}).

We can only tentatively suppose that weak {\MgII} systems with
accompanying {\FeII} absorption may be selecting the class of giant
LSB galaxies (\cite{sprayberry93}) out to $z \sim 1$.
Impey \& Bothun (1989), upon reexamining the selection effects and
assumptions that go into the calculations of galaxy cross sections
from QSO absorbers, found that LSB galaxies are expected to dominate
the absorption cross section.
If we assume a non--evolving absorber cross section, $n_0 \sigma _0$,
we can write the projected radial extent of QSO absorbers as $R \simeq
7.5 (N/n_0)^{1/2}h_{75}^{-1}$~kpc, where we have integrated over the
redshift interval $0.42 \leq z \leq 1.18$, and where $N$ is the number
of observed absorbers in the redshift interval.
If we are to claim that the two {\MgII} systems arise from the general
population of LSB galaxies, we obtain $R\sim 60$~kpc, where we have
used the space density, $n=0.03$~galaxies~$h_{75}^3$~Mpc$^{-3}$, found
by Dalcanton \etal (1997).
Supposing that giant LSB galaxies comprise 1\% of the LSB
population, the size of the absorbers is inferred to be $R\sim
200$~kpc, which compares to the {\HI} sizes of F568--6, UGC~6614, and
UGC~5709 (the ``Malin cousins'').

The success rate of two weak {\MgII} absorbers out of 28 {\Lya} clouds
leads us to tentatively suggest that 5--10\% of the so--called {\Lya}
clouds in the forest at $\left< z \right> \sim 0.7$ will have
detectable {\MgII} absorption to our level of sensitivity.
Using the $dN/dz$ for the {\Lya} forest from the Quasar Absorption
Line Key Project (\cite{buell}), this corresponds to non--evolving
population with $dN/dz \sim 1.5-3$.
Interestingly, this is not inconsistent with the number density of
$dN/dz \simeq 1.74$ for {\MgII} absorbers with $W_{\rm r} \leq
0.3$~{\AA}, as reported by Churchill \etal (1997\nocite{crcv97}).
As with the {\Lya} absorbers at $z\leq1$, the weak {\MgII} systems are
consistent with a non--evolving population.
It could be that our search through the {\qso} forest has picked up
the same population of absorbers selected by a fair
fraction of the weak {\MgII} survey, whatever that population may be.
{\FeII} absorption is present in many of these systems, suggesting
that many of these absorbers may have chemical and photoionization
conditions similar to the two absorbers along the {\qso} sight line
(i.e.~they may arise in similar environments with similar evolutionary
histories).

In contrast to post--starbursting dwarfs, which have not formally been
ruled out if they both happen to be tightly aligned on the sky with
the QSO, the few giant LSB galaxies known to date have colors
consistent with a population of late--type stars (\cite{quillen2}).
This is consistent with the inferred ionization conditions of the two
absorbers, since early--type stars are ruled out as sources of UV flux.
Most giant LSB galaxies have active galactic nuclei [narrow emission
lines (\cite{sprayberry95}; Bothun \etal 1997\nocite{bothun})], and
this could very well be the case here.
There is a point--like/stellar object (\#7 in LBBD) with
impact parameter $\sim 45$ or $60h_{75}^{-1}$ (assuming  $z=0.6$ or
$z=0.9$.) that has a one--sided spiral arm like structure.
Though many damped {\Lya} absorbers are seen to be LSB galaxies and low
luminosity dwarfs (\cite{cohen}; \cite{lebrundla}; \cite{ccs3c336};
\cite{meyer}), this is not a necessary condition for the absorbers to be
LSB galaxies; the high redshift LSB absorbing galaxies studied so far
have been selected by their {\HI} absorption and not by their {\MgII}
absorption.
Weak {\MgII} absorption that is accompanied by {\FeII} absorption of
comparable strength may be selecting a well--defined population of
luminous objects.
These objects probably do not include the damped {\Lya} systems, but
probably do include sub--Lyman limit systems with high metallicity.
This is in contrast to the findings of BBLD, who find that
strong {\MgII} absorption accompanied by strong {\FeII} absorption
selects damped {\Lya} systems of low metallicity.

The weak {\MgII} absorbers may in fact be selecting LSB galaxies,
given that the weak {\MgII} absorbers found by Churchill \etal
(1997\nocite{crcv97}) do not have HSB candidates to $\leq
0.06~L^{*}_{K}$ (\cite{steideleso}; \cite{chuckprivcomm}). 
Very deep imaging and faint object spectroscopy will be required if we
are to identify the luminous objects selected by weak {\MgII}
absorption.
These objects may represent a significant fraction of the galaxy
population of the universe (and therefore dark matter content).
Thus, understanding their statistical properties is important for
theories of structure formation and galactic evolution.

In the future, a detailed comparison of the {\it relative\/}
abundances of $\alpha$--group elements (O, Ne, Mg, Si, S, Ca) to
Fe--peak elements (Cr, Ni, Fe, Zn) in ``{\Lya} forest clouds'' holds
the promise of revealing their various origins and formation epochs.
The rate of Type~Ia SNe appears to be very low for the first Gyr
in the history of a Milky--Way like galaxy (\cite{truran};
\cite{smecker}).
If the delay is similar for the onset of Type Ia SNe in other galaxies
as well, then it could be that [Mg/Fe] abundance ratio tests are
confined to low redshifts.
As such, it is expected that $\alpha$--group enhanced abundance ratios
should be almost exclusively seen at $z > 1.5$ (Timmes \etal
1995\nocite{fxt}).
For $q_0 = 0.5$ and $\Lambda = 0$, if a galaxy formed at $z \geq
4$, then Type Ia SNe may  not {\it begin\/} to contribute to its
chemical enrichment until $z=1.5$ and may not influence the abundance
patterns to a detectable level until $z\sim1$ (1 Gyr later).
For $q_0 = 0.1$ and $\Lambda = 0$, the Fe--group enrichment would
likely be seen no earlier than $z\sim 1.5$.

%%%%%%%%%%%%%%%%%%%%%%%%%%% SECTION %%%%%%%%%%%%%%%%%%%%%%%%%%%%%%%%%%%%%%%%%%%

\section{Summary}
\label{sec:conclude}

We searched for {\MgIIwaves} absorption in 28 {\Lya} forest absorbers
along the {\qso} line of sight.  
The spectrum studied for the metal line absorption was an optical
HIRES spectrum (\cite{mythesis}; Churchill \etal
1998\nocite{cvc98}).
The {\Lya} line list were taken from the UV G190H and the G270H FOS/HST
spectra of BBLD.
The redshift range was $0.4163 \leq z(\lambda 2796) \leq 1.1871$.
The doublets were identified and confirmed using the techniques
described in Schneider \etal (1993) and in Churchill \etal
(1997\nocite{crcv97}).
We found two {\MgII} absorbers, one at $z=0.6428$, and one at
$z=0.9315$.
Both these systems exhibit {\FeII} absorption.
We carefully searched the FOS/HST spectrum for the expected
metal--line transitions (cf.~\cite{uffe}) that were covered in the
forest.
There are currently no other high spectral resolution studies
of metals in the $z<1$ {\Lya} forest, and it also appears that there
are no counter parts to these two systems at higher redshifts.

In Figure~\ref{fig:ewlimits}, we present the 5$\sigma$ rest--frame
observed equivalent width detection limit of the {\MgII} $\lambda
2796$ transition as a function of redshift.
The detection limit ranged from
0.007 to 0.020~{\AA}, except for $z(\lambda 2796) \leq 0.4662$, where
it ranges from 0.020 to 0.035~{\AA}.
The 5$\sigma$ mean upper limit is $\log N({\MgII}) \sim 11.3$~{\cm2}
for clouds with $0.1 \leq W_{\rm r}({\Lya}) \leq 1.6$~{\AA}, which
corresponds to neutral hydrogen column densities over the range $13.5
\leq \log N({\HI}) \leq 18.5$~{\cm2}.

We studied the two discovered absorbers in some detail.
Voigt profile fitting was performed to obtain the column densities and
Doppler $b$ parameters of the {\MgII} and {\FeII}.
Since the {\MgII} and {\FeII} are unresolved, we performed Monte Carlo
modeling of the Voigt profile fitting in order to best constrain the
uncertainties in the Voigt profile parameters (\cite{mythesis}; this
work).
We then used CLOUDY (\cite{ferland}) to model the ionization and
chemical conditions of the two absorbers. 
CLOUDY was used in its optimize mode, in which the residuals between
the model and the measured {\MgII} and {\FeII} column densities were
minimized.
The fixed quantities for each cloud, which constitute the grid
parameters, were the UV flux intensity and continuum shape, the metal
abundance pattern, and the ``observed'' neutral hydrogen column
density.
The optimized output quantities were the total hydrogen density,
$n_H$, and a scaling factor for the metallicity, $Z_{\rm scale}$.
We constrained the model clouds using the {\HI} column density and the
parameter $f=b_{\rm turb}/b_{\rm tot}$, where $b_{\rm turb}$ is the
non--thermal contribution to $b_{\rm tot}$.

The UV continua considered were a Haardt \& Madau (1996) UV background
(UVB) spectrum, several Kurucz (1991) Atlas stellar models, late--type
galaxy models (\cite{worthey}), and a star forming galaxy model
(\cite{bruzual}).
We investigated three abundance patterns: solar, {\HII} depletion, and
ISM depletion (the latter being $\alpha$--group enhanced).  
The {\HII} and ISM CLOUDY models included grain physics.
The only UV ionizing scenario that yielded model clouds that were
consistent both with astrophysical constraints (numbers of stars, etc.)
and with constraints imposed by the data was the Haardt \& Madau UVB.
We conclude that the absorbers are photoionized by the UVB and not
by stellar radiation.
Neither absorber is consistent with having an $\alpha$--group enhanced
abundance pattern.

As described in \S\ref{sec:0.6428}, the $z=0.6428$ absorber may have a
near--solar or super--solar [Fe/H].
For the solar abundance pattern, the model cloud has
$16.3 \leq \log N({\HI}) \leq 16.8$~{\cm2}, a $b_{\rm turb}/b_{\rm tot}$
of $0.85 \leq f \leq 0.93$, metallicity $-0.2 \geq
[Z/Z_{\odot}] \geq -0.7$, and density $0.01 \leq n_{\rm H} \leq
0.02$~cm$^{-3}$.
For the {\HII} abundance pattern, the absorber has $16.7 \leq \log
N({\HI}) \leq 17.2$~{\cm2}, $b_{\rm turb}/b_{\rm tot}$ in the range
$0.90 \leq f \leq 0.95$, metallicity $+0.4 \geq [Z/Z_{\HII}]
\geq 0.0$, and density $n_{\rm H} \simeq 0.008$~cm$^{-3}$.
If this cloud is relatively free of dust depletion, so that the
abundance pattern is close to solar, then the cloud has $[\hbox{Fe/H}]
> -1$.
If the gas--phase abundance follows that of depleted clouds in our
Galaxy, then the cloud could have $[\hbox{Fe/H}] > 0$.

As described in \S\ref{sec:0.9315}, the $z=0.9315$ model cloud appears
to have a super--solar gas--phase [Fe/H].
For the solar abundance pattern, the absorber has $15.8 \leq \log
N({\HI}) \leq 16.3$~{\cm2}, a $b_{\rm turb}/b_{\rm tot}$ of $0.62 \leq
f \leq 0.94$, metallicity $+0.7 \geq [Z/Z_{\odot}] \geq +0.1$,
and density $0.2 \leq n_{\rm H} \leq 0.4$~cm$^{-3}$.
The inferred $[\hbox{Fe/H}]$ is even greater for the {\HII} abundance
pattern model.
For the {\HII} abundance pattern, the absorber has $16.0 \leq \log
N({\HI}) \leq 16.5$~{\cm2}, $b_{\rm turb}/b_{\rm tot}$ in the range
$0.75 \leq f \leq 0.97$, metallicity $+1.6 \geq [Z/Z_{\HII}]
\geq +0.7$, and density $n_{\rm H} \simeq 0.1$~cm$^{-3}$.
This is a metallicity enhancement of five to 40 times over the typical
values seen in Galactic {\HII} regions (\cite{baldwin91};
\cite{rubin91}; Osterbrock \etal 1992\nocite{osterbrock92}).   
No matter the abundance pattern, this cloud has $[\hbox{Fe/H}] > 0$.

In \S\ref{sec:discuss}, we discussed the possibility that these two
absorbers could arise in giant LSB galaxies.
These galaxies are seen to have metallicities that scatter about solar
(\cite{pickering95}) and large extended disks (\cite{quillen}).
We tentatively suggest that 5\% (at most 10\%) of $z \leq 1$ ``{\Lya}
forest clouds'' with  $0.1 \leq W_{\rm r}({\Lya}) \leq 1.6$~{\AA} will
exhibit {\MgII} absorption to a 5$\sigma$ $W_{\rm r}$ detection limit
of 0.02~{\AA}.
The sub--sample of these systems that also exhibit comparable {\FeII}
absorption may have iron--group enhanced metallicities with
$[\hbox{Fe/H}] \geq -1$, and may be selecting giant LSB galaxies at
high redshifts.

%%%%%%%%%%%%%%%%%%%%%%%%%%% ACKNOWLEDGMENTS %%%%%%%%%%%%%%%%%%%%%%%%%%%%%%%%%%%

\acknowledgments
This work has been supported in part by the National Science
Foundation grant AST--9617185 at Penn State.
CWC acknowledges support through the Eberly School of Science
Distinguished Postdoctoral Fellowship at The Pennsylvania State
University.
Thanks to:
P.~Boiss\'e for providing the FOS spectrum prior to publication;
J.~Charlton for assistance with the Voigt profile fitting simulations
and CLOUDY modeling; G.~Ferland for making CLOUDY a public tool,
U.~Hellsten for providing the Haardt \& Madau input spectra for CLOUDY;
S.~Linder for generating plots of $\mu_{B}(0)$ versus $\log
\alpha$ of LSB galaxies for our inspection and for discussions about
the observed properties of LSB galaxies; and 
C.~Steidel for sharing unpublished data on the {\qso} field.
It is a pleasure to acknowledge J.~Bergeron, M.~Bershady, P.~Boiss\'e,
J.~Charlton, R.~Dav\'e, U.~Hellsten, C.~Impey, J.~Lauroesch,
P.~Petitjean, D.~Schneider, and M.~Shetrone for stimulating
discussions and/or comments.
Special thanks to S.~Vogt for HIRES and for providing the opportunity
to use it.
We thank C. Impey, the referee, for valuable comments that improved
the quality of this manuscript.
For J.~L.~N.

%%%%%%%%%%%%%%%%%%%%%%%%%%% APPENDIX %%%%%%%%%%%%%%%%%%%%%%%%%%%%%%%%%%%%%%%%%%

\section*{APPENDIX}

\appendix

\section{Searching the FOS Spectrum}
\label{app:fossearch}

It is important to thoroughly check for the presence of absorption from
from both low and high ionization species in the FOS spectrum.
Here, we present our search for absorption in the FOS spectrum from
both the $z=0.6428$ and $z=0.9315$ systems.
We then obtained the limiting column densities (to 3.5$\sigma$) using
curve of growth analysis, where we have assumed both pure thermal and
pure turbulent scaling to the measured {\MgII} $b$ parameters.
These limits may be useful for further
constraining the chemical and ionization conditions of the absorber,
even if the sensitivity level in the FOS spectrum is not very high.
The continuum fit has been adopted from the work of BLLD.
We have used interactive software of our own that employed the line
detection algorithm described in Schneider \etal (1993).
Since these transitions are embedded in the {\Lya} forest, it is very
difficult to make identifications at a high confidence level.
Selected results are tabulated in Table~\ref{tab:limits}.

\subsection{The $z=0.6428$ Absorber}

A 3.5$\sigma$ observed equivalent width detection limit,
which corresponds to approximately 0.25~{\AA}.
Based upon preliminary CLOUDY models of this system, we have used the
Line Observability Index (LOX) of Hellsten \etal (1997) to place
limits on those transitions that would most likely be detectable in
the spectrum.  
Roughly in decreasing order of their LOX, the covered transitions
include  {\CIV} $\lambda \lambda 1548, 1550$, 
{\SiIII} $\lambda 1206$, 
{\SiIV} $\lambda \lambda 1394, 1402$, 
{\CII} $\lambda 1335$ and $\lambda 1036$, 
{\SIV} $\lambda 1063$, 
{\SIII} $\lambda 1190$, 
{\OVI} $\lambda \lambda 1032, 1038$,  
{\SiII} $\lambda 1260$, and
{\SII} $\lambda 1259$.

Both the expected positions of the {\SiIII} $\lambda 1206$ and {\SIV}
$\lambda 1063$ transitions are coincident with the wings of higher
order Lyman series lines from the strong $z=1.1538$ {\MgII} absorption
system.
There is no detected {\SIII} $\lambda 1190$, which suggests that the
{\SIV} is likely not present in absorption.
Both the {\SiII} $\lambda 1260$ and {\SII} $\lambda 1259$ transition 
are not detected, nor are the {\SiII} $\lambda 1190$, 1304, and 1527
transitions. 
Thus, it appears that {\SiII}, {\SiIII}, {\SII}, and {\SIII} are not
detected to a 3.5$\sigma$ observed limit of 0.25~{\AA}, which
corresponds to the upper limits $\log N \sim 15.4$, 15.3, 18.7, and
16.1~{\cm2}, respectively.
Additionally, neither the {\AlII} $\lambda 1671$ nor the {\AlIII}
$\lambda 1855$ transitions are detected, giving limits $\log N
\sim 14.3$ and 14.5~{\cm2}, respectively.
The QSO flux is extinct due to the Lyman limit break of the $z=0.8596$
damped system at the expected positions of {\CII} $\lambda 1036$ and
the {\OVI} doublet.
However, the {\CII} $\lambda 1335$ transition is not detected, which
provides a column density limit of $\log N \sim 15.2$~{\cm2} for
{\CII}.
We also obtained (not very useful) upper limits on {\NII} $\lambda
1084$ and {\FeIII} $\lambda 1122$, giving $\log N \sim 16.7$ and
17.6~{\cm2}.

Determining the limits on the {\CIV} and {\SiIV} doublets is less clear.
The {\CIV} $\lambda 1548$ transition, predicted to be observed at
2543.37~{\AA}, lies precisely at the position of a $z=1.0922$ {\Lya}
line with observed equivalent width 0.64~{\AA}, as identified by
BBLD.
There are no other transitions, including {\Lyb}, that corroborate
the {\Lya} absorber.
We have also measured the wavelength and observed equivalent width of
the feature and obtain $\lambda_{\rm obs} = 2543.27\pm0.14$~{\AA} and
$W_{\rm obs} = 0.57\pm0.09$~{\AA}.
There is an unresolved 2.5$\sigma$ absorption feature 0.51~{\AA}
redward ($+90$~{\kms}) of the expected {\CIV} $\lambda 1550$
transition with $\lambda_{\rm obs} = 2548.11\pm0.18$~{\AA} and $W_{\rm
obs} = 0.22\pm0.08$~{\AA}.
Though the doublet ratio is consistent with physically allowed values,
there are at least two reasons to not adopt this feature
as the {\CIV} $\lambda 1550$ line at $z=0.6428$. 
First, the detection falls below a 3.5$\sigma$ significance when the
continuum fit is adjusted downward by only 0.5$(S/N)^{-1}$, where
$S/N$ is the signal to noise in the continuum of the normalized
spectrum.  
The continuum placement is not obvious to this level of accuracy over
the region $2540 - 2558$~{\AA}, which brackets the putative {\CIV}
line.
Second, the measured $\lambda_{\rm obs}$ is approximately a 3$\sigma
_{\lambda}$ difference from the predicted $\lambda_{\rm obs}$.

The region of the FOS spectrum where the {\SiIV} doublet transitions
are predicted to fall (2289.66 and 2304.47~{\AA}, respectively)
exhibits what appears to be a low level flux depression.
The only feature identified by BBLD is a broad and
asymmetric $z=0.8816$ {\Lya} line.
The {\SiIV} $\lambda 1393$ transition could be a blend in the red wing
of this {\Lya} line; the optimized equivalent width spectrum (following
Schneider \etal 1993) shows that there are two minima, suggestive of
a two component blend with $\lambda _{\rm obs} = 2286.4$ and
2289.5~{\AA}.
The latter is consistent with the predicted location of the $\lambda
1393$ transition.
Additionally, the optimized equivalent width spectrum has a single
minimum at 2304.7~{\AA}, consistent with the predicted location of the
$\lambda 1402$ transition.
These minima have $\sim 3.5\sigma$ detection significance levels.
However, the continuum fit is also somewhat uncertain over this region
of the spectrum, so that it is difficult to assess if the {\SiIV}
doublet has actually been observed in absorption.
If the {\SiIV} is present, then the {\CIV} is most certainly present.

All things considered, we adopt the position that the {\CIV} doublet
has not been detected in the FOS spectrum and assign the 3.5$\sigma$
observed equivalent width limit of 0.25~{\AA} to the {\CIV} $\lambda
1550$ transition, the $\lambda 1548$ transition being compromised by
the $z=1.0922$ {\Lya} line.
This gives an upper limit column density of $\log N \sim 14.7$~{\cm2}
for {\CIV}. 
We also adopt the position that the {\SiIV} doublet has not been
detected, which gives the limiting column density $\log \sim
15.7$~{\cm2}.
However, these are not secure conclusions.  
If the absorber is multiphased, then our analysis is affected only
in the sense that our single--phase cloud model must be valid only for
the cooler phase with the higher density.
If so, then the estimated {\HI} column density is an upper limit in
this phase of the cloud, since some fraction of it must then arise in
the hotter, lower density phase.

\subsection{The $z=0.9315$ Absorber}

Again, the 3.5$\sigma$ observed equivalent width limit of the spectrum
is approximately 0.25~{\AA}, and we have used the LOX of Hellsten
\etal (1997) to place limits on those transitions that would most
likely be detectable in the spectrum.  
In approximate order of their LOX, the covered transitions include 
{\CIII} $\lambda 977$, 
{\CIV} $\lambda \lambda 1548, 1550$, 
{\SiIII} $\lambda 1206$,
{\SiIV} $\lambda \lambda 1394, 1403$, 
{\NIII} $\lambda 990$, 
{\CII} $\lambda 1335$ and $\lambda 1036$, 
{\SIII} $\lambda 1012$ and $\lambda 1190$, 
{\SIV} $\lambda 1063$,
{\OVI} $\lambda \lambda 1032, 1038$, and
{\SiII} $\lambda 1260$, among others. 

None of these transitions were detected to the 3.5$\sigma$ level.
For the clear non detections, the corresponding upper limit column
densities are 
$\log N({\CIII}) \sim 16.3$~{\cm2}, 
$\log N({\SiIV}) \sim 16.1$~{\cm2}, 
$\log N({\NIII}) \sim 17.9$~{\cm2},
$\log N({\SIII}) \sim 16.2$~{\cm2}, and
$\log N({\SiII}) \sim 15.6$~{\cm2}.
For a few of the transitions, the limits are affected by the presence
of other absorption features in the spectrum.

The continuum level fit in the region where the {\SIV} $\lambda 1063$
transition is predicted is somewhat uncertain, but if the BBLD
fit is adopted, then the not very stringent upper limit is
$\log N({\SIV}) \sim 17.9$~{\cm2}.
The {\CIV} doublet, if present, would be blended in both transitions.
The $\lambda 1548$ component is coincident with the blue wing of the
{\FeII} $\lambda 1608$ transition of the $z=0.8598$ damped {\Lya}
absorber and the $\lambda 1550$ component is coincident with the blue
wing of the {\MgII} $\lambda 2796$ transition of the $z=0.0714$ dwarf
galaxy (see \cite{ccsdwarf}; \cite{lebrundla}).
Thus, we cannot place useful limits on the {\CIV} column density.

The expected position of the {\SiIII} $\lambda 1206$ transition is
coincident with a $z=0.8598$ {\OVI} $\lambda 1038$ line from the 
damped {\Lya} absorber.
Otherwise the limit would be $\log N({\SiIII}) \sim 15.5$~{\cm2}.
The $\lambda 1206$ transition is the only transition with which limits
on {\SiIII} could have been placed.
There is clearly no absorption feature where the {\CII}
$\lambda 1036$ transition is expected, giving the not very stringent
upper limit on {\CII} of $\log N \sim 16.8$~{\cm2}.
The expected position of the {\OVI} $\lambda 1038$ transition is
coincident with a higher order {\HI} line from the strong {\MgII}
absorption system at $z=1.1537$.
However, a non--stringent  upper limit of $\log N \sim 18.6$~{\cm2} is
obtained for {\OVI} via the $\lambda 1032$ transition, which would
arise in a featureless region of the spectrum.

Interestingly, there is a formally significant absorption feature at
$\lambda _{\rm obs} = 2168.16$~{\AA} with $W_{\rm obs} = 0.37\pm0.09$
that is coincident with {\FeIII} $\lambda 1122$ at $z=0.9315$. 
Assuming a thermal line broadening in a single--phase cloud, this
implies the very large column density of $\log N \sim 17.8$~{\cm2}.
This large of a column density is virtually impossible to reconcile
with the {\FeII} measurement.
There are two alternative possible identifications for the putative
{\FeIII} line.
First, it could be a {\Lya} line at $z=0.7835$, though there is no
corroborating evidence to support this interpretation.
Second, it could be {\Lyb} at $z=1.1137$, which would be supported by
the presence of a {\Lya} line at $\lambda _{\rm obs} = 2569.58$~{\AA}.
BBLD detected an asymmetric line with an extended blue
wing at 2569.85~{\AA}, which they identified as a blend of {\NV} $\lambda
1242$ at $z=1.0678$ (associated with a {\CIV} system) with {\SiII}
$\lambda 1193$ at $z=1.1536$ (from the strong {\MgII} system).
The {\NV} is the weaker transition of the $\lambda \lambda 1238, 1242$
doublet and the {\SiII} is the stronger transition of the $\lambda
\lambda 1190, 1193$ doublet.
The {\NV} $\lambda 1238$ and {\SiII} $\lambda 1190$ components of these
doublets have been identified by BBLD in two adjacent and
unambiguous absorption features, so the blend is corroborated by 
the doublet counterparts.
If both these features have been correctly identified and the
2569.85~{\AA} feature is a blend of the doublets, then there appears
to be an inconsistency in that unphysical doublet ratios are required.
The equivalent width of the {\NV} $\lambda 1238$ -- {\SiII} $\lambda
1190$ blend has a lower limit 0.2~{\AA} greater than the measured
equivalent width (under the assumption of optically thick {\SiII} and
optically thin {\NV}).
Also, the identified blend is asymmetric in such a way as to suggest
its centroid may in fact be 2569.6~{\AA} (the predicted position of
the {\Lya} line if the line in question at 2168.16~{\AA} were {\Lyb}).
If the blend were actually a {\Lya} line, it would place the {\NV}
$\lambda 1238$ and/or the {\SiII} $\lambda 1190$ identifications in
question because the observed {\Lya} equivalent width would need to
account for at least 50\% of the absorption.
This particular discussion serves to illustrate the level of ambiguity
that arises when line identifications in the {\Lya} forest are being
considered.

We adopt the position that the line observed at 2168.16~{\AA} is
{\it not\/} the {\FeIII} $\lambda 1122$ transition at $z=0.9315$. 
The strongest argument is that the LOX of the {\FeIII} $\lambda 1122$
transition is significantly smaller than that of {\CIII} $\lambda
977$, which has an ionization potential a bit higher than {\FeIII} and
is therefore likely to be present in an environment giving rise to
{\FeIII}.
If {\FeIII} were present, the {\CIII} transition  would have likely
been easily seen in absorption.

\section{Computing the Number of Stars}
\label{app:numstars}

Assuming no interstellar extinction, stars of all spectral types and
luminosity classes produce an observed $\nu f_{\nu} = \lambda
f_{\lambda} = 10^{-4.7}$~ergs~{\cm2}~s$^{-1}$ at $\lambda =
5500$~{\AA} when their apparent visual magnitudes are $V=0$
(\cite{allen}).
Integrating over space out to some finite radius, $R$, which contains
a constant density of stars with a given spectral type, one obtains
\begin{equation}
\log \left[ \nu f_{\nu}(0.17) \right] _{N_{\ast}} =   \log N_{\ast} -
2\log R + C_{\ast} ,
\label{eq:Nstars}
\end{equation}
where $R$ is in kiloparsecs, and $\left[ \nu f_{\nu}(0.17) \right]
_{N_{\ast}}$ is the total flux from $N_{\ast}$ stars at $\lambda =
5500$~{\AA}, which corresponds to 0.17~Ryd.
The value of $C_{\ast}$, given by $\log 3 \nu f_{\nu}(0.17) + 2\log
d_{\ast} - 6$, is dependent upon the distance, $d_{\ast}$~[pc], at
which the given stellar type has apparent visual magnitudes $V=0$.
If the stars are not located in a constant density sphere centered on
the clouds, but instead are all at some distance, $R$, the only
modification to eq.~[{\ref{eq:Nstars}}] is that each $C_{\ast}$ is
reduced by a geometric factor, $\log 3 \sim 0.5$.

The $C_{\ast}$ are dependent upon the stellar luminosity class, I,
III, and V.  For O5, B0, A0, and G0 stars, the $C_{\ast}$ are 
\begin{equation}
\begin{array}{lcl}
C_{\rm O5}\hbox{(I,V)}&=&(-5.9,-5.9) , \\
C_{\rm B0}\hbox{(I,III,V)}&=&(-5.8,-6.2,-6.6) , \\ 
C_{\rm A0}\hbox{(I,III,V)}&=&(-5.8,-7.9,-8.5) , \\
C_{\rm G0}\hbox{(I,III,V)}&=&(-5.8,-8.6,-10.5) .
\end{array}
\end{equation}
Note that all supergiant stars, luminosity class I, have roughly
identical $C_{\ast}$.

We are most interested in the flux intensity at $1-1.2$~Ryd, since
this is the energy range corresponding to the ionization potentials of
the observed transitions.
To estimate the number of stars required to elevate the integrated
stellar/galactic flux such that it equals the UVB $\nu f_{\nu}$ at
$\sim 1.2$~Ryd, we need to introduce a continuum shape term into
eq.~[\ref{eq:Nstars}].
We define $\log k_{\ast}$ to be the ratio of the stellar flux at
0.17~Ryd to that at 1.2~Ryd, or $\log \left[ \nu f_{\nu}(0.17) \right]
_{N_{\ast}} -  \log \left[ \nu f_{\nu}(1.20) \right] _{N_{\ast}}$.
For a given metallicity, Atlas stellar models (\cite{kurucz}) show
that the $\log k_{\ast}$ are not sensitive to surface gravity (luminosity
class), but only to effective surface temperature (spectral type).
Thus, eq.~[\ref{eq:Nstars}] can be written,
\begin{equation}
\log N_{\ast} - 2\log R =  \log k_{\ast} + \log \left[ \nu
f_{\nu}(1.20) \right] _{\rm UVB} - C_{\ast} ,
\label{eq:Nabove}
\end{equation}
where $\log \left[ \nu f_{\nu}(1.20) \right] _{N_{\ast}}$ has been
replaced by $\log \left[ \nu f_{\nu}(1.20) \right] _{\rm UVB}$ (under
the assumption of no shielding or extinction of the UVB), which ranges
from $10^{-5.2} \leq \left[ \nu f_{\nu}(1.20) \right] _{\rm UVB} \leq
10^{-5.6}$~ergs~{\cm2}~s$^{-1}$ in the redshift range of interest (see
Figure~\ref{fig:uvb}).

The spectral shapes, especially continuum breaks, play a significant
role in the number of stars that are required to dominate, or match,
the UVB flux at $\sim 1$~Ryd.
For O, B, A, and G stars, respectively, the $\log k_{\ast}$ are
roughly 0, 3, 9, and $\infty$.
Thus, for a cloud embedded in a $R=1$~kpc volume of space, the number
of luminosity class V stars required to influence the ionization
conditions of the model clouds are $N_{\ast} \sim 1, 10^{4}, 10^{12}$,
and $\infty$ stars, respectively.
For supergaints, the numbers are $N_{\ast} \sim 1, 10^{3}, 10^{9}$, and
$\infty$ stars, respectively.
At redshifts of $0.5 \leq z \leq 1.0$, a single O star could carve out 
a $R=1$~kpc volume in which it could match or exceed the UVB at
$1-1.2$~Ryd.
However, it would require 1000 B0I stars, or 10,000 B0V stars
distributed in that same volume.
For luminosity class III stars, the numbers are $N_{\ast} \sim
10^{4}, 10^{11}$, and $\infty$ stars, respectively, where the O type
star has been omitted.
Because of their sharp {\HI} break, G, K, and M stars would need to
have astrophysically unrealistic number densities to contribute  to
the ionization conditions of the absorbers, regardless of their
luminosity class.

It is of interest to know if the implied stellar number densities are
consistent with those of the Galaxy.
In terms of the stellar number density per cubic parsec, $n_{\ast}$,
eq.~[\ref{eq:Nabove}] is written,
\begin{equation}
\log n_{\ast} + \log R = -9.6 + \log k_{\ast} + \log \left[ \nu
f_{\nu}(1.20) \right] _{\rm UVB} - C_{\ast} , \quad \hbox{stars
pc$^{-3}$}
\label{eq:starden}
\end{equation}
where $R$, as above, is in kpc.
In a volume with $R=1$~kpc, the required number density of O stars,
whether main sequence or supergiant, is $n_{\rm O} \sim
10^{-9}$~stars~pc$^{-3}$.
For the remaining spectral types, the required stellar density is
dependent upon their luminosity class.  
For an A0 star, the required number densities are $n_{\rm A}
\sim 1, 100$, and 100~stars~pc$^{-3}$ for luminosity classes I, III,
and V, respectively.
In contrast, the observed stellar density of all giants in the solar
neighborhood is vanishingly small, and for A0V stars is
$10^{-3.3}$~stars~pc$^{-3}$ (\cite{allen}).

Clearly, under the assumption that the UVB is not shielded by layers
of neutral hydrogen or extinguished due to dust, the type of
stellar environment that would be required for the stellar flux to
dominate over the UVB at $\sim 1$~Ryd would either need to be
uncommonly dense or populated by O stars.
Lower metallicity stars have smaller breaks and flatter overall
continua; lower metallicity stars have smaller $\log k_{\ast}$.
Thus, the solar metallicity stars provide a rough upper limit to the
number of required stars.
On the other hand, interstellar extinction would have the effect of
reducing $\nu f_{\nu}$ at 1.2~Ryd for the individual stars, which
makes the above estimates a lower limit for solar metallicity stars.

%%%%%%%%%%%%%%%%%%%%%%%%%%% REFERENCES %%%%%%%%%%%%%%%%%%%%%%%%%%%%%%%%%%%%%%%%

%%%%%%%%%%%%%%%%%%%%%%%%%%% FIGURE CAPTIONS %%%%%%%%%%%%%%%%%%%%%%%%%%%%%%%%%%%

%%%%%%%%%%%%%%%%%%%%%%%%%%% TABLES %%%%%%%%%%%%%%%%%%%%%%%%%%%%%%%%%%%%%%%%%%%%

\include{tab1pp}
\include{tab2}
\include{tab3}
\include{tab4}
\include{tab5}

%%%%%%%%%%%%%%%%%%%%%%%%%%% FIGURES %%%%%%%%%%%%%%%%%%%%%%%%%%%%%%%%%%%%%%%%%%%

\begin{figure}[th]
\plotone{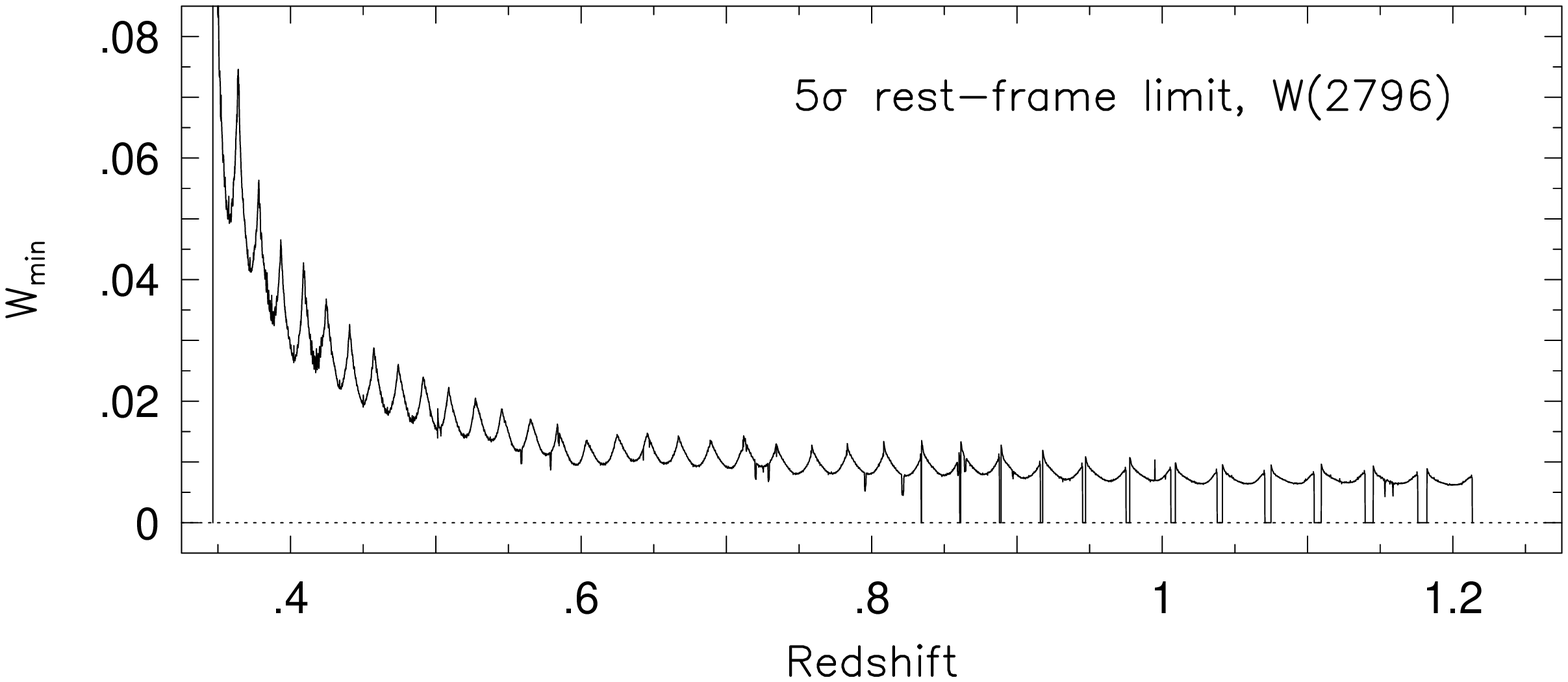}
\figurenum{1}
\caption{The 5$\sigma$ rest--frame equivalent width detection limit
of the {\MgII} $\lambda 2796$ transition as a function of redshift.
The decrease in sensitivity toward lower redshift is due to the
HIRES throughput and CCD efficiencies.  The higher frequency features
are due to the blaze efficiency function of the individual echelle
orders.  Gaps in the redshift coverage are appear at $z \sim 0.83$ and
above.}
\label{fig:ewlimits}
\end{figure}

\begin{figure}[th]
\plotone{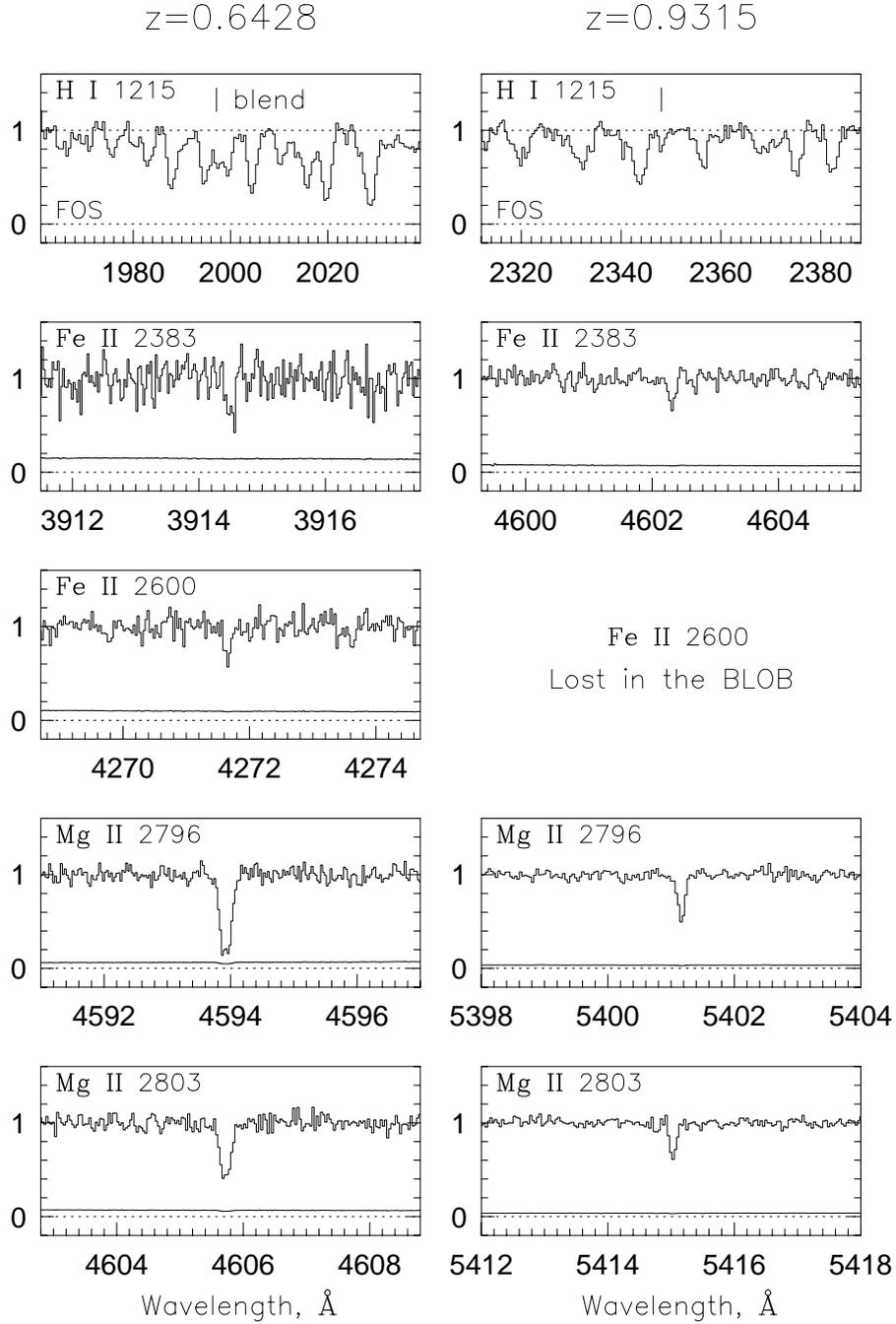}
\figurenum{2}
\caption{The FOS/HST (top panels) and HIRES/Keck (metal
lines) spectra of the two absorbers.  Note the very different
wavelength scales presented between the FOS and the HIRES spectra.
The ticks mark the predicted positions of {\Lya} $\lambda$1215
absorption lines based upon the accurate redshifts of the metal lines.
The $z=0.6428$ {\Lya} line is a member of a blended feature (see
Fig.~\ref{fig:midas}).} 
\label{fig:data}
\end{figure}

\begin{figure}[th]
\plotone{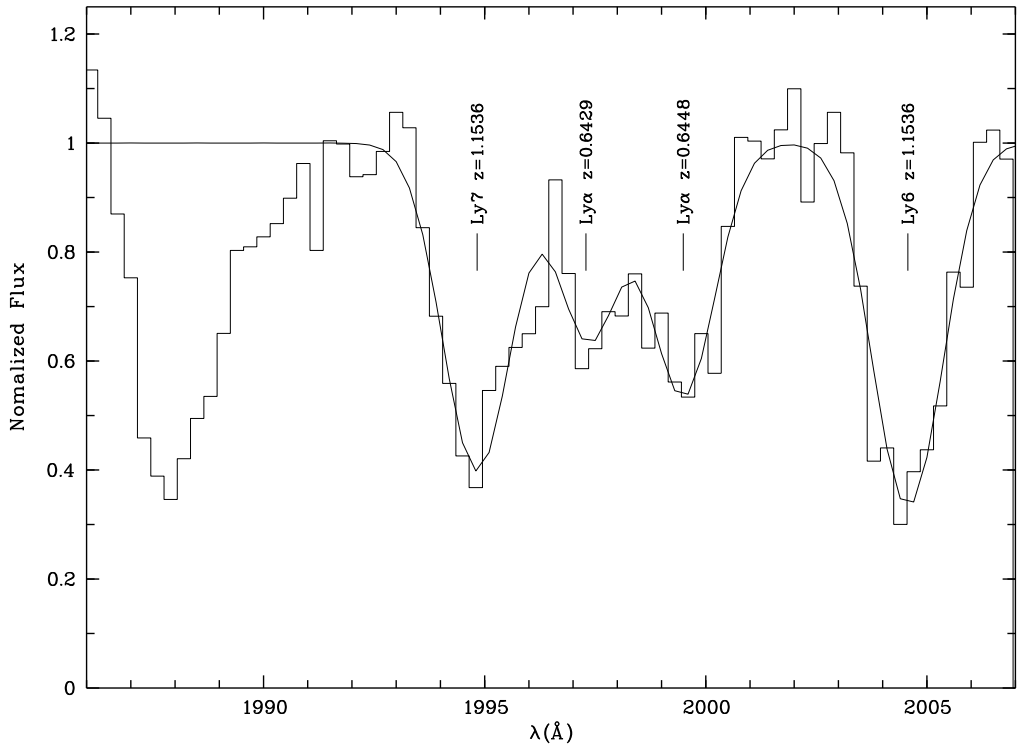}
\figurenum{3}
\caption{The deblending fit for the $z=0.6428$ {\Lya}
line (second from the left) in the FOS spectrum.  The blended
lines are {\Lyseven} at $z=1.1536$, {\Lya} at $z = 0.6429$, and
{\Lya} at $z=0.6448$.  The lone feature at 2004.5~{\AA} is {\Lysix} at
$z=1.1536$, which is the redshift of a strong {\MgII} absorber.}
\label{fig:midas}
\end{figure}

\begin{figure}[th]
\plotone{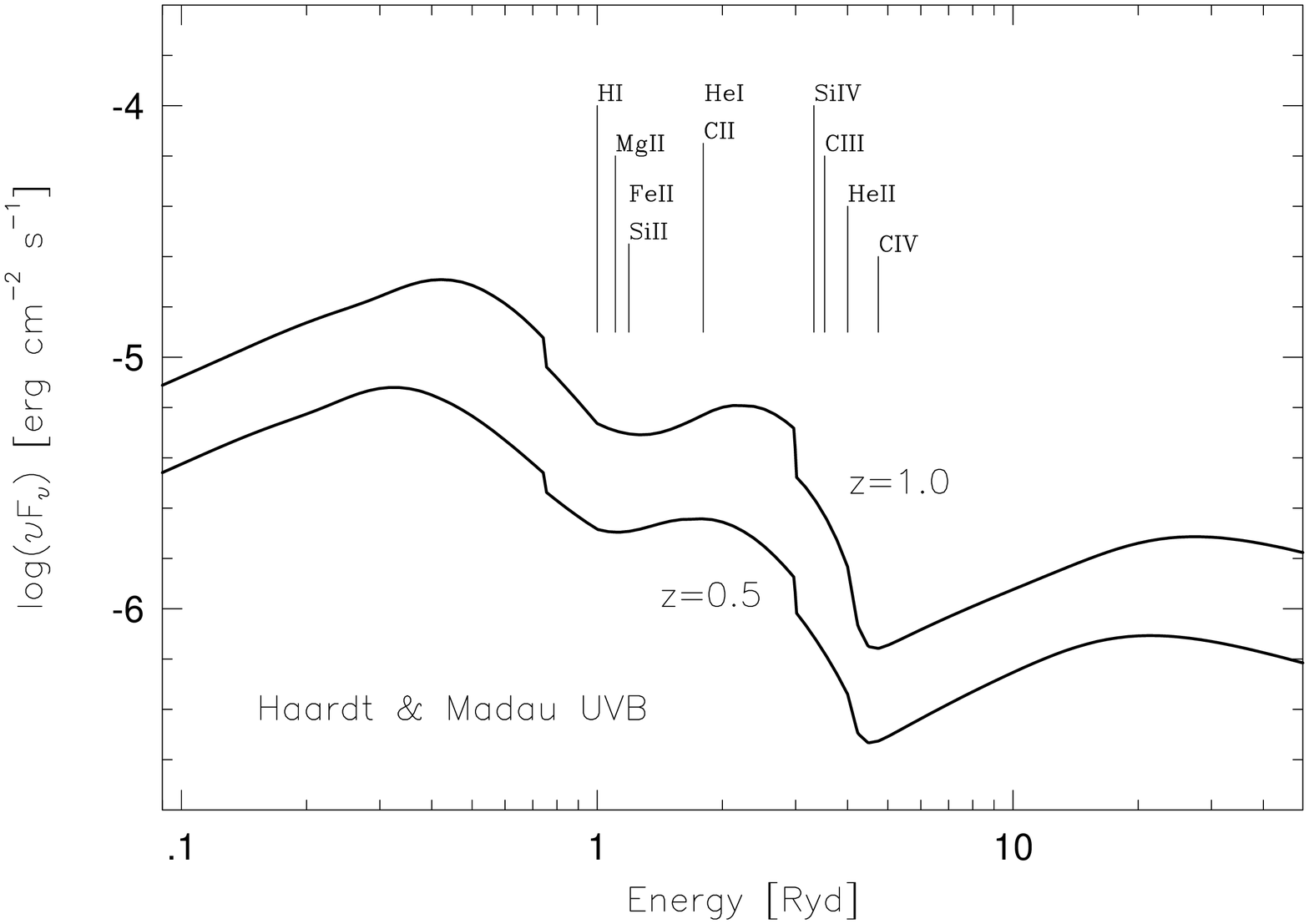}
\figurenum{4}
\caption{The Haardt and Madau UVB normalization and shape
for redshifts 1.0 and 0.5, as labeled.  The ionization potentials for
selected ion species are shown as vertical ticks in the range $E
\geq 1$~Ryd.}
\label{fig:uvb}
\end{figure}

\begin{figure}[th]
\plotone{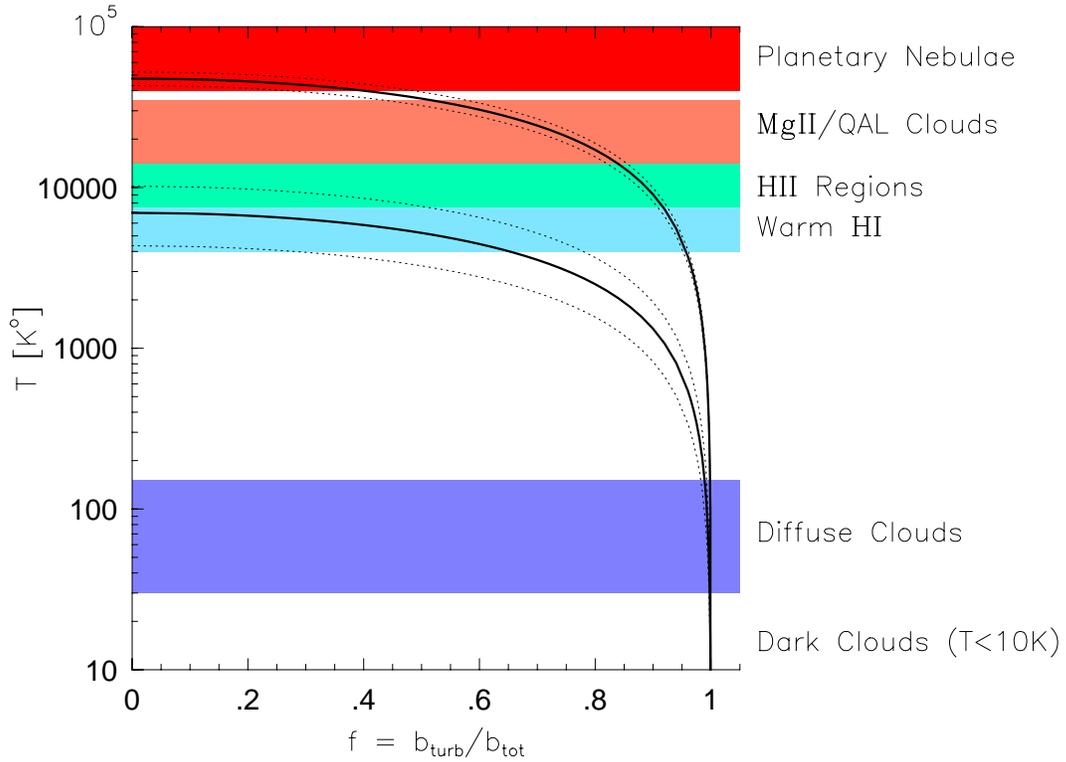}
\figurenum{5}
\caption{The inferred temperature of the $z=0.6428$ absorber
(upper curves) and of the $z=0.9315$ absorber (lower curves) as a
function of the fraction of the measured {\MgII} $b$ parameter due to
possible turbulence, $f=b_{\rm turb}/b_{\rm tot}$.  The uncertainties
in $T$, based upon the $1\sigma$ uncertainties in the {\MgII}
$b_{\rm tot}$, are given by the dotted curves.  The shaded regions
give the typical range of $T$ for various objects in the Galactic
ISM and for {\MgII} QSO Absorption Line ({\MgII}/QAL) clouds at $z\sim1$.
The observed absorbers are consistent with {\MgII}/QAL clouds, {\HII}
Regions, and Warm {\HI} clouds (see text).}
\label{fig:temp}
\end{figure}

\begin{figure}[th]
\plotone{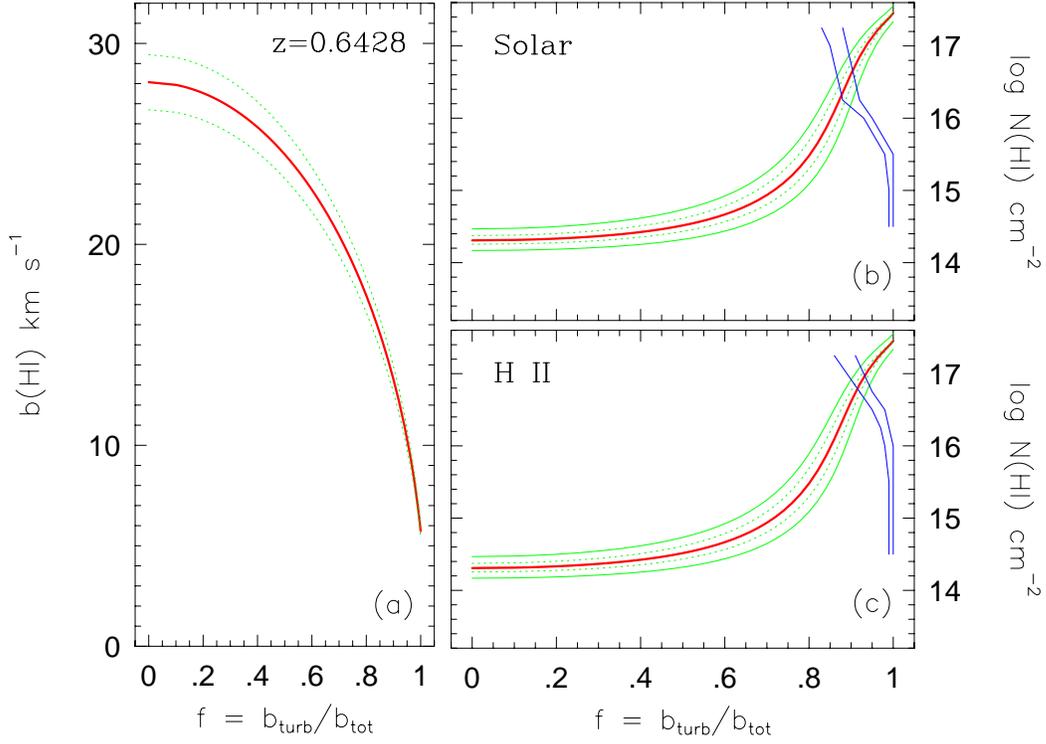}
\figurenum{6}
\figcaption{The $z=0.6428$ absorber properties 
--- (a) The range of total Doppler $b$ parameters as a function of
$f=b_{\rm turb}/b_{\rm tot}$. The dotted lines are the spread due the
$1\sigma$ uncertainties in the measured {\MgII} $b_{\rm tot}$. 
--- (b) and (c) The range of {\HI} column densities as a function
of $f$.  The thick solid curves are best values of
$N({\HI})$ based upon $b_{\rm tot}({\MgII})$ and $W_{\rm
r}({\HI})$.  The thin curves give the spread of $N({\HI})$ based upon
the $1\sigma$ uncertainties in $b_{\rm tot}({\MgII})$ [inner] and
$W_{\rm r}({\Lya})$ [outer].  The curves that originate in the lower
right hand corners and rise upward and then to the left are the
allowed locus of $f$ for a cloud model with a given $N({\HI})$.
Panel (b) is for a solar abundance patterns and (c) is for the {\HII}
abundance pattern.}
\label{fig:z06428}
\end{figure}

\begin{figure}[th]
\plotone{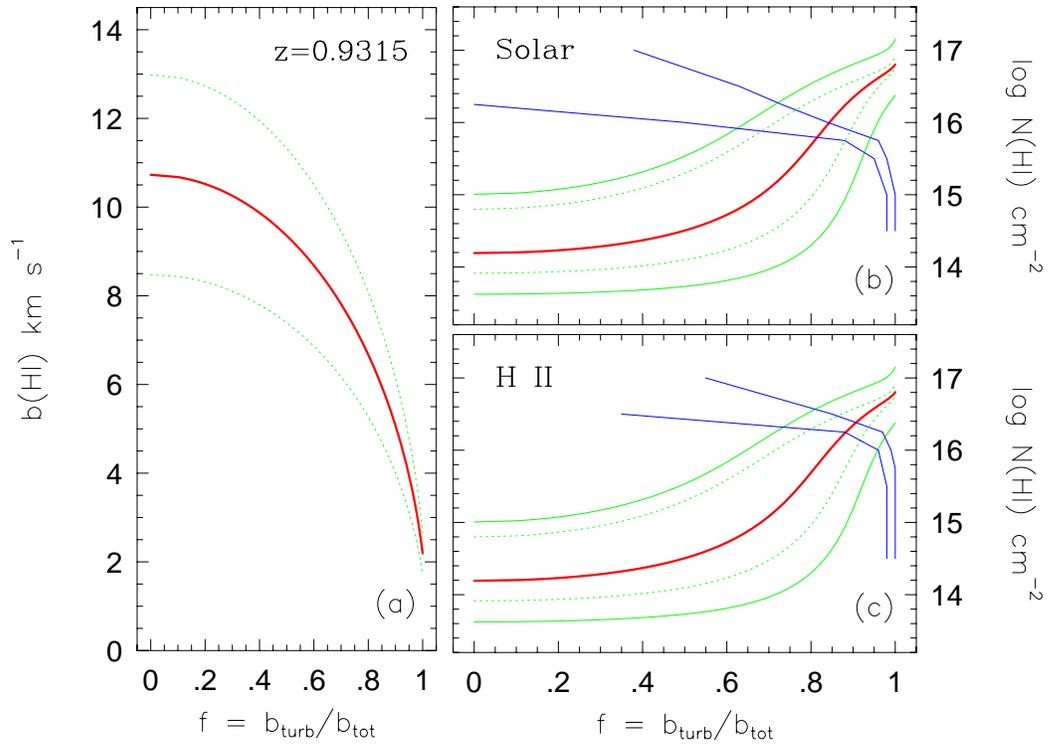}
\figurenum{7}
\figcaption{The $z=0.9315$ absorber properties.  See caption for
Figure~\ref{fig:z06428}.}
\label{fig:z09315}
\end{figure}

\end{document}

%% file: tab1pp.tex
\begin{deluxetable}{ccccccccc}
\tablenum{1}
\tablewidth{0pc}
\tablecolumns{9}
\tablecaption{Equivalent Width Limits in the {\Lya} Forest}
\tablehead
{
\multicolumn{2}{c}{Boiss\'e et~al.~1997} &
\colhead{ } &
\multicolumn{3}{c}{{\HI}} &
\colhead{ } &
\multicolumn{2}{c}{{\MgII}} \\
\cline{1-2} \cline{4-6} \cline{8-9}
\colhead{$z_{\rm abs}({\Lya})$} &
\colhead{$W_{\rm r}({\Lya})$} & 
\colhead{ } &
\colhead{$\log N_{80}$} &  
\colhead{$\log N_{30}$} &
\colhead{$\log N_{15}$} & 
\colhead{ } &
\colhead{$W_{\rm r,lim}$} &
\colhead{$\log N$} \\
\colhead{ }  &
\colhead{[{\AA}]} &
\colhead{ }  &
\colhead{[{\cm2}]} &  
\colhead{[{\cm2}]} &  
\colhead{[{\cm2}]} &  
\colhead{ }  &
\colhead{[{\AA}]} &
\colhead{[{\cm2}]}   
}
\startdata
0.4163 & 1.03 & & 14.9 & 18.1 & 18.5 & & 0.025 & 11.8 \nl
0.4662 & 0.37 & & 14.0 & 14.4 & 16.7 & & 0.017 & 11.6 \nl
0.5577 & 0.36 & & 14.0 & 14.4 & 16.5 & & 0.013 & 11.5 \nl
0.5898 & 0.63 & & 14.3 & 16.3 & 17.8 & & 0.013 & 11.5 \nl
0.6251 & 0.35 & & 13.9 & 14.3 & 16.4 & & 0.012 & 11.4 \nl
0.6428\tablenotemark{a} & 0.33 & & 13.9 & 14.3 & 16.4 & & 0.012 & 11.4 \nl
0.6447 & 0.75 & & 14.5 & 17.2 & 18.1 & & 0.015 & 11.6 \nl
0.6786 & 0.28 & & 13.8 & 14.0 & 15.4 & & 0.009 & 11.3 \nl
0.6968 & 0.19 & & 13.6 & 13.7 & 14.2 & & 0.009 & 11.3 \nl
0.7566 & 0.29 & & 13.8 & 14.1 & 15.5 & & 0.012 & 11.4 \nl
0.7934 & 0.11 & & 13.3 & 13.4 & 13.5 & & 0.009 & 11.3 \nl
0.7986 & 0.11 & & 13.3 & 13.4 & 13.5 & & 0.009 & 11.3 \nl
0.8069 & 0.39 & & 14.0 & 14.5 & 16.9 & & 0.011 & 11.4 \nl
0.8125 & 0.35 & & 13.9 & 14.3 & 16.4 & & 0.011 & 11.4 \nl
0.8816 & 0.53 & & 14.2 & 15.4 & 17.6 & & 0.008 & 11.3 \nl
0.9084 & 0.59 & & 14.3 & 15.9 & 17.8 & & 0.008 & 11.3 \nl
0.9183 & 0.78 & & 14.5 & 17.3 & 18.2 & & 0.011 & 11.4 \nl
0.9315\tablenotemark{a} & 0.15 & & 13.5 & 13.6 & 13.9 & & 0.008 & 11.3 \nl
0.9384 & 0.51 & & 14.2 & 15.3 & 17.6 & & 0.008 & 11.3 \nl
0.9781 & 1.63 & & 16.5 & 18.8 & 19.2 & & 0.008 & 11.3 \nl
0.9948 & 1.19 & & 15.3 & 18.4 & 18.7 & & 0.008 & 11.3 \nl
1.0558 & 0.50 & & 14.1 & 15.2 & 17.5 & & 0.008 & 11.2 \nl
1.0679 & 0.41 & & 14.0 & 14.6 & 17.0 & & 0.008 & 11.2 \nl
1.0922 & 0.31 & & 13.9 & 14.2 & 15.8 & & 0.007 & 11.2 \nl
1.0995 & 0.40 & & 14.0 & 14.6 & 17.0 & & 0.007 & 11.2 \nl
1.1674 & 0.86 & & 14.6 & 17.7 & 18.3 & & 0.007 & 11.2 \nl
1.1749 & 0.31 & & 13.9 & 14.2 & 15.8 & & 0.007 & 11.2 \nl
1.1871 & 0.16 & & 13.5 & 13.6 & 13.9 & & 0.007 & 11.2 \nl
\enddata
\tablenotetext{a}{These two systems were detected in {\MgII}, but are
presented here for purpose of illustration.}
\tablecomments{Columns 1 and 2 give the redshift and rest--frame {\Lya} 
equivalent width as measured by Boiss\'e et~al.\ (1997), except where
noted.  Columns 3, 4, and 5 give the estimated neutral hydrogen
column density for $b({\HI}) = 80, 30$, and 15~{\kms}, respectively.
Columns 6 and 7 give the 5$\sigma$ rest--frame {\MgII}~$\lambda 2796$
equivalent width limit and {\MgII} column density assuming linear
curve of growth analysis for the $W_{\rm r,lim}({\MgII})$.}
\label{tab:mgiilimits}
\end{deluxetable}

%% file: tab2.tex
\begin{deluxetable}{lcccc}
\tablenum{2}
\tablewidth{0pc}
\tablecolumns{5}
\tablecaption{Measured Properties of Absorbing Systems}
\tablehead
{
\colhead{Transition} &
\colhead{$\lambda _{\rm obs}$} &
\colhead{$W_{\rm r}$} &
\colhead{$\log N$} &
\colhead{$b$} \\
\colhead{ } &
\colhead{[{\AA}]} &
\colhead{[{\AA}]} &
\colhead{[{\cm2}]} &  
\colhead{[{\kms}]} 
}
\startdata
\cutinhead{$z=0.6428$ Absorber}
{\HI} 1216   & 1997.10  & $0.33\pm0.03$   &                &               \nl
{\FeII} 2383 & 3914.504 & $0.031\pm0.012$ & $12.46\pm0.06$ & $4.29\pm1.25$ \nl
{\FeII} 2600 & 4271.654 & $0.029\pm0.004$ &                &               \nl
{\MgII} 2796 & 4593.924 & $0.120\pm0.008$ & $12.74\pm0.02$ & $5.73\pm0.28$ \nl
{\MgII} 2803 & 4605.719 & $0.086\pm0.004$ &                &               \nl
\cutinhead{$z=0.9315$ Absorber}  
{\HI} 1216   & 2348.07  & $0.15\pm0.04$   &                &               \nl
{\FeII} 2383 & 4602.312 & $0.020\pm0.003$ & $12.24\pm0.09$ & $2.28\pm1.57$ \nl
{\MgII} 2796 & 5401.148 & $0.041\pm0.002$ & $12.29\pm0.08$ & $2.19\pm0.46$ \nl
{\MgII} 2803\tablenotemark{a} & 5415.035 & $0.020\pm0.001$ &                &               \nl
\enddata
\tablenotetext{a}{Based upon the Monte Carlo simulations, the
equivalent width of this transition may actually be 0.023~{\AA}.}
\tablecomments{Columns 4 and 5 are based upon Voigt Profile fits to
the HIRES data.  The tabulated column densities and $b$ parameters
apply to all transitions of a given ion species; however, they are given
opposite only to the strongest transition for the species.}
\label{tab:systems}
\end{deluxetable}

%% file: tab3.tex
\begin{deluxetable}{lcccccc}
\tablenum{3}
\tablewidth{0pc}
\tablecolumns{7}
\tablecaption{FOS Limiting Column Densities of {\MgII} Absorbers}
\tablehead
{
\colhead{ } &
\colhead{ } &
\multicolumn{2}{c}{Thermal Limits} &
\colhead{ } &
\multicolumn{2}{c}{Turbulent Limits} \\
\cline{3-4} \cline{6-7}
\colhead{Transition} &
\colhead{$\lambda _{\rm exp}$} &
\colhead{$b$} &
\colhead{$\log N$} &  
\colhead{} &
\colhead{$b$} &
\colhead{$\log N$} \\
\colhead{ } &
\colhead{[{\AA}]} &
\colhead{[{\kms}]} &
\colhead{[{\cm2}]} &  
\colhead{} &
\colhead{[{\kms}]} &
\colhead{[{\cm2}]} 
}
\startdata
\cutinhead{$z=0.6428$ Absorber}
{\SIII} $\lambda 1190$                 & 1955.65 & 4.9 & $ <16.1$ & & 5.7 & $<15.9$ \nl
{\SiIII} $\lambda 1206$                & 1982.07 & 5.3 & $ <15.3$ & & 5.7 & $<15.2$ \nl
{\SiII} $\lambda 1260$                 & 2070.66 & 5.3 & $ <15.4$ & & 5.7 & $<15.3$ \nl
{\CII} $\lambda 1335$                  & 2194.34 & 8.1 & $ <15.2$ & & 5.7 & $<16.4$ \nl
{\SiIV} $\lambda \lambda 1394, 1402$\tablenotemark{a} & 2289.70, 2304.51 & 5.3 & $ <15.6$ & & 5.7 & $<15.2$ \nl
{\CIV} $\lambda \lambda 1548, 1550$\tablenotemark{a}  & 2543.42, 2574,65 & 8.1 & $ <14.7$ & & 5.7 & $<15.7$ \nl
\cutinhead{$z=0.9315$ Absorber}  
{\CIII} $\lambda 977$                  & 1887.11 & 3.1 & $<16.3$ & & 2.2 & $<16.4$ \nl
{\OVI} $\lambda \lambda 1032, 1038$    & 1993.17, 2004.15 & 2.7 & $<18.6$ & & 2.2 & $<18.7$ \nl
{\CII} $\lambda 1036$                  & 2001.68 & 3.1 & $<16.8$ & & 2.2 & $<17.0$ \nl
{\SIII} $\lambda 1190$,                & 2299.29 & 2.0 & $<16.2$ & & 2.2 & $<16.2$ \nl
{\SiIII} $\lambda 1206$                & 2330.35 & 2.0 & $<15.5$ & & 2.2 & $<15.5$ \nl
{\SiII} $\lambda 1260$                 & 2434.51 & 2.0 & $<15.6$ & & 2.2 & $<15.6$ \nl
{\SiIV} $\lambda \lambda 1394, 1403$   & 2692.04, 2709.45 & 2.0 & $<16.1$ & & 2.2 & $<16.1$ \nl
{\CIV} $\lambda \lambda 1548, 1550$    & 2990.34, 2995.31 & 2.0 & $<16.7$ & & 2.2 & $<16.8$ \nl
\enddata
\tablenotetext{a}{These limits on these transitions are not highly
certain (see Appendix~\ref{app:fossearch}).}
\label{tab:limits}
\end{deluxetable}

%% file: tab4.tex
\def\gp{\phantom{-}}

\begin{deluxetable}{ccccccccl}
\tablenum{4}
%\tablewidth{450pt}
\tablewidth{0pc}
\tablecolumns{9}
\tablecaption{Haardt \& Madau UVB: $z=0.6428$ Optimized CLOUDY Models}
\tablehead
{
\colhead{$Z_{\rm scale}$\tablenotemark{a}} & 
\colhead{$\Delta ({\MgII})$} & 
\colhead{$\Delta ({\FeII})$} & 
\colhead{$\log n_{\rm H}$} &  
\colhead{$N({\HI})$} & 
\colhead{$N({\HII})$} & 
\colhead{Temp} &
\colhead{$b_{\rm turb}/b_{tot}$} &
\colhead{Note\tablenotemark{b}} \\
\colhead{ } &
\colhead{[dex]} & 
\colhead{[dex]} & 
\colhead{[{\cm2}]} &  
\colhead{[{\cm2}]} & 
\colhead{[{\cm2}]} & 
\colhead{[$^{\circ}$K]} &
\colhead{ } &
\colhead{ }
}
\startdata
\cutinhead{Solar Abundance Pattern; No Grains}
       $\gp2.0$ & $\gp0.03$ & $\gp0.00$ & $-2.4$ & 14.50 & 15.2 &   170 & $0.99-1.00$ & N \nl
       $\gp0.9$ & $\gp0.04$ & $\gp0.00$ & $-1.9$ & 15.50 & 16.3 &   720 & $0.98-1.00$ & N \nl
       $\gp0.1$ & $\gp0.03$ & $\gp0.00$ & $-2.0$ & 16.00 & 17.4 &  6310 & $0.93-0.94$ & N \nl 
       $  -0.2$ & $  -0.01$ & $\gp0.00$ & $-2.0$ & 16.25 & 17.8 &  7940 & $0.88-0.92$ & Y \nl
       $  -0.4$ & $\gp0.03$ & $\gp0.00$ & $-2.0$ & 16.50 & 18.0 & 10000 & $0.87-0.90$ & Y \nl
       $  -0.6$ & $\gp0.00$ & $\gp0.00$ & $-1.7$ & 16.75 & 18.1 & 10200 & $0.86-0.89$ & Y \nl
       $  -0.8$ & $\gp0.03$ & $\gp0.00$ & $-1.6$ & 17.00 & 18.3 & 11200 & $0.85-0.88$ & Y: \nl
       $  -1.0$ & $\gp0.00$ & $\gp0.00$ & $-1.5$ & 17.25 & 18.4 & 11500 & $0.84-0.88$ & N \nl
\cutinhead{{\HII} Abundance Pattern; Grains}
       $\gp3.0$ & $\gp0.00$ & $\gp0.00$ & $-2.9$ & 14.50 & 15.3 &    90 & $0.99-1.00$ & N \nl
       $\gp2.0$ & $\gp0.00$ & $\gp0.01$ & $-2.4$ & 15.50 & 16.4 &   200 & $0.99-1.00$ & N \nl
       $\gp1.5$ & $\gp0.00$ & $  -0.01$ & $-2.2$ & 16.00 & 16.9 &   490 & $0.98-1.00$ & N \nl
       $\gp1.2$ & $\gp0.00$ & $\gp0.00$ & $-2.1$ & 16.25 & 17.3 &  1170 & $0.97-1.00$ & N \nl   
       $\gp0.7$ & $\gp0.01$ & $\gp0.00$ & $-2.1$ & 16.50 & 18.0 &  4170 & $0.95-0.96$ & N: \nl
       $\gp0.4$ & $\gp0.00$ & $\gp0.00$ & $-2.1$ & 16.75 & 18.3 &  6310 & $0.92-0.94$ & Y \nl      
       $\gp0.2$ & $\gp0.00$ & $\gp0.00$ & $-2.1$ & 17.00 & 18.6 &  8130 & $0.90-0.92$ & Y \nl
       $  -0.1$ & $\gp0.00$ & $\gp0.00$ & $-2.1$ & 17.25 & 18.8 & 10000 & $0.88-0.90$ & N: \nl  
\enddata
\tablenotetext{a}{Optimal logarithmic scaling factor applied to all
       elements heavier than helium for the given abundance patterns.}
\tablenotetext{b}{Notes: (Y=yes, N=no) indicate whether CLOUDY
       conditions are consistent with inferred $f=b_{\rm
       turb}/b_{tot}$ and allowed values of $N({\HI})$ and $b_{\rm
       tot}$ shown in Figure~\ref{fig:z06428}. A ``:'' indicates a
       borderline case.}
\label{tab:hm06428}
\end{deluxetable}

%% file: tab5.tex
\def\gp{\phantom{-}}

\begin{deluxetable}{ccccccccl}
\tablenum{5}
%\tablewidth{450pt}
\tablewidth{0pc}
\tablecolumns{9}
\tablecaption{Haardt \& Madau UVB: $z=0.9315$ Optimized CLOUDY Models}
\tablehead
{
\colhead{$Z_{\rm scale}$\tablenotemark{a}} & 
\colhead{$\Delta ({\MgII})$} & 
\colhead{$\Delta ({\FeII})$} & 
\colhead{$\log n_{\rm H}$} &  
\colhead{$N({\HI})$} & 
\colhead{$N({\HII})$} & 
\colhead{Temp} &
\colhead{$b_{\rm turb}/b_{tot}$} &
\colhead{Note\tablenotemark{b}} \\
\colhead{ } &
\colhead{[dex]} & 
\colhead{[dex]} & 
\colhead{[{\cm2}]} &  
\colhead{[{\cm2}]} & 
\colhead{[{\cm2}]} & 
\colhead{[$^{\circ}$K]} &
\colhead{ } &
\colhead{ }
}
\startdata
\cutinhead{Solar Abundance Pattern; No Grains}
       $\gp2.0$ & $\gp0.03$ & $  -0.03$ & $-1.1$ & 14.50 & 14.5 &  100 & $0.98-1.00$ & N \nl
       $\gp0.9$ & $\gp0.04$ & $  -0.04$ & $-0.8$ & 15.50 & 15.6 &  420 & $0.95-0.98$ & N \nl
       $\gp0.7$ & $\gp0.04$ & $  -0.03$ & $-0.6$ & 15.75 & 15.9 &  810 & $0.90-0.96$ & Y \nl      
       $\gp0.3$ & $\gp0.03$ & $  -0.03$ & $-0.6$ & 16.00 & 16.3 & 3310 & $0.50-0.82$ & Y \nl 
       $\gp0.1$ & $\gp0.03$ & $  -0.04$ & $-0.4$ & 16.25 & 16.6 & 4790 & $0.00-0.73$ & Y \nl  
       $  -0.1$ & $\gp0.03$ & $  -0.03$ & $-0.2$ & 16.50 & 16.7 & 5890 & $0.00-0.63$ & N \nl
       $  -0.7$ & $\gp0.03$ & $  -0.03$ & $-0.2$ & 17.00 & 17.3 & 8910 & $0.00-0.38$ & N \nl
\cutinhead{{\HII} Abundance Pattern; Grains}
       $\gp2.9$ & $\gp0.00$ & $\gp0.00$ & $-1.8$ & 14.50 & 14.8 &   80 & $0.98-1.00$ & N \nl
       $\gp2.0$ & $\gp0.00$ & $\gp0.00$ & $-1.2$ & 15.50 & 15.6 &  110 & $0.98-1.00$ & N \nl
       $\gp1.8$ & $\gp0.00$ & $\gp0.00$ & $-1.1$ & 15.75 & 15.8 &  130 & $0.97-1.00$ & N \nl      
       $\gp1.6$ & $\gp0.00$ & $\gp0.00$ & $-0.9$ & 16.00 & 16.0 &  200 & $0.96-0.99$ & Y \nl
       $\gp1.3$ & $\gp0.01$ & $\gp0.01$ & $-0.8$ & 16.25 & 16.3 &  440 & $0.94-0.97$ & Y \nl   
       $\gp0.7$ & $\gp0.01$ & $\gp0.00$ & $-0.8$ & 16.50 & 17.1 & 3890 & $0.35-0.79$ & Y \nl
       $\gp0.2$ & $\gp0.00$ & $\gp0.00$ & $-0.6$ & 17.00 & 17.6 & 6920 & $0.00-0.55$ & N \nl
\enddata
\tablenotetext{a}{Optimal logarithmic scaling factor applied to all
       elements heavier than helium for the given abundance patterns.}
\tablenotetext{b}{Notes: (Y=yes, N=no) indicate whether CLOUDY
       conditions are consistent with inferred $f=b_{\rm
       turb}/b_{tot}$ and allowed values of $N({\HI})$ and $b_{\rm
       tot}$ shown in Figure~\ref{fig:z09315}.}
\label{tab:hm09314}
\end{deluxetable}